\documentclass{emulateapj}

\def\hdoneoh{\object[HD 108317]{HD~108317}}
\def\hdonetwo{\object[HD 122563]{HD~122563}}
\def\hdonefour{\object[HD 140283]{HD~140283}}

\def\bd{\object[BD+44 493]{BD$+$44~493}}

\def\kmsec{\mbox{km~s$^{\rm -1}$}}
\def\logg{\mbox{log~{\it g}}}

\def\teff{\mbox{$T_{\rm eff}$}}
\def\vt{\mbox{$v_{\rm t}$}}
\def\logrw{\mbox{$\log(W_{\lambda}/\lambda)$}}

\def\loggf{$\log gf$}

\shorttitle{Phosphorus abundances in metal-poor stars}
\shortauthors{Roederer et al.}
\slugcomment{Accepted for publication in the Astrophysical Journal}

\begin{document}

\title{Detection of Neutral Phosphorus in the
Near Ultra-violet Spectra \\ of Late-Type Stars\footnotemark[1]}

\footnotetext[1]{
Based on observations made with the NASA/ESA 
\textit{Hubble Space Telescope},
obtained at the Space Telescope Science Institute, 
which is operated by the Association of Universities for Research in 
Astronomy, Inc., under NASA contract NAS~5-26555. 
This work is supported through program AR-13246 and is based
on observations associated with programs
GO-7348,
GO-7433,
GO-8197,
GO-9048,
GO-9049,
GO-9455,
GO-9804, 
GO-12268,
GO-12554, and
GO-12976.
Portions of this work are
based on data obtained from the European Southern Observatory (ESO) 
Science Archive Facility.
These data are associated with Programs
065.L-0507(A),
067.D-0439(A),
072.B-0179(A),
074.C-0364(A),
076.B-0055(A), and
266.D-5655(A).~
Portions of this research have also 
made use of the Keck Observatory Archive (KOA), 
which is operated by the W.M.\ Keck Observatory and the NASA Exoplanet 
Science Institute (NExScI), under contract with the 
National Aeronautics and Space Administration.
These data are associated with Programs
H2aH (P.I.\ Boesgaard),
H5aH (P.I.\ Stephens), and
H47aH (P.I.\ Boesgaard).
Other portions of this work are based on data gathered with the 6.5~meter 
Magellan Telescopes located at Las Campanas Observatory, Chile, and
the McDonald Observatory of The 
University of Texas at Austin.
}

\author{
Ian U.\ Roederer,\altaffilmark{2}
Heather R.\ Jacobson,\altaffilmark{3}
Thanawuth Thanathibodee,\altaffilmark{3}
Anna Frebel,\altaffilmark{3}
Elizabeth Toller\altaffilmark{3}
}

\altaffiltext{2}{Department of Astronomy, University of Michigan,
1085 S.\ University Ave., 
Ann Arbor, MI 48109, USA; iur@umich.edu
}
\altaffiltext{3}{Department of Physics \&
Kavli Institute for Astrophysics and Space Research, 
Massachusetts Institute of Technology, 
77 Massachusetts Ave., 
Cambridge, MA 02139, USA
}


\addtocounter{footnote}{3}

\begin{abstract}

We report the detection of several 
absorption lines of neutral phosphorus (P, $Z =$~15)
in archival near-ultraviolet spectra 
obtained with the 
Space Telescope Imaging Spectrograph 
on board the \textit{Hubble Space Telescope}.
We derive phosphorus abundances or interesting upper limits
in 14 late-type stars with 
metallicities spanning $-$3.8~$<$~[Fe/H]~$< -$0.1.
Previously, phosphorus had only been studied in 
Galactic stars with $-$1.0~$<$~[Fe/H]~$< +$0.3.
Iron lines reveal abundance
offsets between the optical and ultraviolet
regions, and we discuss and apply a correction
factor to account for this offset.
In stars with [Fe/H]~$> -$1.0, 
the [P/Fe] ratio decreases toward the solar value
with increasing metallicity,
in agreement with previous observational studies.
In stars with [Fe/H]~$< -$1.0, 
$\langle$[P/Fe]$\rangle = +$0.04~$\pm$~0.10,
which overlaps with the [P/Fe] ratios found in 
several high-redshift damped Lyman-$\alpha$ systems.
This behavior 
hints at
a primary origin in massive stars.

\end{abstract}


\section{Introduction}
\label{intro}

Phosphorus (P, $Z =$~15) is one of the few
light elements 
whose nucleosynthetic origins in ancient stars
remain unexamined by observations.
Models by
\citet{woosley95} and \citet{kobayashi06}
predict the phosphorus yields of massive supernovae.
These yields are incorporated into chemical evolution models, 
like those of \citet{timmes95} and \citet{cescutti12}, 
to predict the Galactic evolution of phosphorus
from the first stars until today.
At present, however, no observations
are capable of verifying these predictions
in low-metallicity Galactic environments.

One neutron-rich isotope of phosphorus, $^{31}$P, is found
in solar system material.
Previous derivations of the phosphorus abundance in the Sun
show remarkable agreement among different studies.
The earliest solar photospheric estimates
by \citet{goldberg60} and \citet{lambert68}
derived 
$\log \epsilon$~(P)~$=$~5.34 and 5.43, 
respectively,
on the scale where $\log \epsilon$~(H)~$\equiv$~12.00.
These and other estimates from the last four decades
are in agreement with modern estimates
by \citet{asplund09} and \citet{caffau07},
5.41 and 5.46, respectively.
These values are also in agreement with the
recommended meteoritic abundance, 5.45
\citep{lodders09}.

\citet{melendez09} derived phosphorus abundances in 
10~solar twins, and 
\citet{caffau11} 
studied the evolution of phosphorus in 20~stars
with $-$1.0~$<$~[Fe/H]~$< +$0.3.
Phosphorus ions have been detected in the atmospheres of 
normal O- and B-type stars (e.g., \citealt{struve30,crowther02}), 
the F-type star Procyon \citep{kato96},
chemically-peculiar A- and B-type stars on the upper main sequence
or horizontal branch
(e.g., \citealt{bidelman60,sargent67,leckrone99,castelli04}), and
hot stars that have evolved beyond the asymptotic giant branch 
(AGB; e.g., \citealt{vennes96,marcolino07,reiff07}).
P-bearing molecules---including CP, PN, PO, HCP, 
C$_{2}$P, and PH$_{3}$---have been detected in 
cold, dense 
clouds (e.g., \citealt{turner87,ziurys87}) 
and the carbon- and oxygen-rich circumstellar envelopes of
stars in the AGB phase of evolution
(e.g., \citealt{guelin90,halfen08,milam08,tenenbaum08}).
P~\textsc{ii} has been detected 
in neutral gas in the interstellar medium (ISM) 
of the Milky Way
(e.g., \citealt{morton75,jenkins86,savage96,lehner03}).
P~\textsc{ii} has been detected in the ISM of
external galaxies, including
the Large Magellanic Cloud \citep{friedman00},
M33 \citep{lebouteiller06}, and
several more distant low-metallicity star-forming galaxies 
(e.g., \citealt{lebouteiller13}).
P~\textsc{ii} has also been detected
in several high-redshift damped Lyman-$\alpha$ systems 
(DLAs; \citealt{outram99,molaro01,lopez02,levshakov02}), 
and
P~\textsc{v} has been detected in 
the broad-absorption-line region of a 
quasar \citep{junkkarinen97}.
The common theme of these extra-solar phosphorus detections
is that they do not offer opportunities
to confront nucleosynthesis predictions
in Galactic environments with metallicity [Fe/H]~$< -$1.0.

In contrast, 
extensive observations are
available for the neighboring elements
silicon (Si, $Z =$~14) and sulfur (S, $Z =$~16).
Observations of
Si~\textsc{i} lines across the optical spectrum 
of late-type stars with $-$4.0~$\leq$~[Fe/H]~$\leq +$0.3
have led to 
consensus regarding the cosmic origins and evolution of this element
(e.g., \citealt{luck81,peterson81,tomkin85,magain87,gratton88,edvardsson93,
ryan96,fulbright00,cayrel04,reddy06,bensby14}).
Sulfur has also been systematically investigated 
using a few multiplets of S~\textsc{i} in the optical and 
near infrared 
(NIR; 
e.g., \citealt{clegg81,francois87,francois88,chen02,
takadahidai02,takadahidai05,reddy03,ecuvillon04,
nissen04,nissen07,
ryde04,caffau05,caffau10,nissen07,jonsson11,
takeda11,spite11,matrozis13}).
Most studies have found that
both [Si/Fe] and [S/Fe] exhibit an enhanced ``plateau'' at low
metallicity.
This echos familiar trends seen in other
elements produced along the $\alpha$ chain, 
most notably oxygen (O, $Z =$~8), 
magnesium (Mg, $Z =$~12), and 
calcium (Ca, $Z =$~20).
Silicon and sulfur are formed primarily
through $\alpha$-captures during hydrostatic and explosive oxygen burning,
and
the enhanced [Si/Fe] and [S/Fe] plateaus
result from nucleosynthesis in core-collapse
supernovae early in the history of the Galaxy
(e.g., \citealt{matteucci89,timmes95,thielemann96,samland98,goswami00,
kobayashi06,tominaga07}).

In stars like the Sun, 
no optical lines of P~\textsc{i}
are available, and only a few weak P~\textsc{i} multiplets
are present in the NIR.~
These NIR lines have been known in the solar spectrum for a long time
\citep{moore34}.  
Only the studies of \citet{kato96},
\citet{melendez09}, and \citet{caffau11}
have derived phosphorus abundances in stars
useful for examining the nucleosynthetic fossil record.
The NIR lines are too weak to detect in
late-type stars with [Fe/H]~$< -$1.0.
Here we show that this observational limitation
can be overcome through the use of 
near ultra-violet (NUV) spectroscopy
of late-type stars.
We examine P~\textsc{i} 
absorption lines present in the near NUV
spectra of late-type stars in the solar neighborhood
spanning $-$3.8~$<$~[Fe/H]~$< -$0.1.
Our results extend the metallicity range
of stars with phosphorus abundances by nearly 3~dex.

We use standard definitions of elemental abundances and ratios.
For element X, the logarithmic abundance is defined
as the number of atoms of element X per 10$^{12}$ hydrogen atoms,
$\log\epsilon$(X)~$\equiv \log_{10}(N_{\rm X}/N_{\rm H}) +12.0$.
For elements X and Y, the logarithmic abundance ratio relative to the
solar ratio of X and Y is defined as
[X/Y]~$\equiv \log_{10} (N_{\rm X}/N_{\rm Y}) -
\log_{10} (N_{\rm X}/N_{\rm Y})_{\odot}$.
Abundances or ratios denoted with the ionization state
indicate the total elemental abundance as derived from 
that particular ionization state.
We adopt the \citet{asplund09} photospheric abundances
for all elements studied.

\section{Observational data}
\label{observations}

\begin{deluxetable*}{ccccccc}
\tablecaption{Characteristics of Archival NUV Spectra
\label{obstab}}
\tablewidth{0pt}
\tabletypesize{\scriptsize}
\tablehead{
\colhead{Star} &
\colhead{$\lambda$} &
\colhead{$R$} &
\colhead{S/N} &
\colhead{Datasets} & 
\colhead{Program} &
\colhead{P.I.} \\
\colhead{} &
\colhead{(\AA)} &
\colhead{$\equiv \lambda/\Delta\lambda$} &
\colhead{2136~\AA} &
\colhead{} &
\colhead{ID} &
\colhead{} 
}
\startdata
BD$+$44~493 &         2280--3070 &  30,000 &   60\tablenotemark{a}
                                                  & QBQ603-04     & GO-12554 & Beers      \\
G64-12    &           2010--2810 &  30,000 &   10 & O6ED01-04     & GO-9049  & Deliyannis \\
HD~2454   &           1970--2810 &  30,000 &   80 & O6E601        & GO-9048  & Deliyannis \\
HD~16220  &           1990--2815 &  30,000 &   90 & O6E603        & GO-9048  & Deliyannis \\
HD~43318  &           1990--2815 &  30,000 &   90 & O6E606        & GO-9048  & Deliyannis \\
HD~76932  &           1880--2148 & 110,000 &   75 & O8P201-02     & GO-9804  & Duncan     \\
HD~94028  &           1885--2147 & 110,000 &   35 & O5CN01-03     & GO-8197  & Duncan     \\
HD~107113 & $\approx$~1800--2360 &  30,000 &  100 & O4AO08        & GO-7433  & Heap       \\
HD~108317 & $\approx$~1900--2365 &  30,000 &  100 & OBXV01-04     & GO-12976 & Roederer   \\
HD~128279 & $\approx$~1900--2365 &  30,000 &   90 & OBXV05-07     & GO-12976 & Roederer   \\
HD~140283 &           1933--2310 & 110,000 &   60 & O55Z01-02     & GO-7348  & Edvardsson \\
          &           2370--3155 & 110,000 &\ldots& O6LM71        & GO-9455  & Peterson   \\
HD~155646 &           1990--2815 &  30,000 &   35 & O6E609        & GO-9048  & Deliyannis \\
HD~160617 &           1880--2147 & 110,000 &   30 & O5CN05-06, 54 & GO-8197  & Duncan     \\
HD~211998 & $\approx$~1900--2147 & 110,000 &   40 & O8P203-04     & GO-9804  & Duncan     
\enddata
\tablenotetext{a}{S/N at 2553~\AA}
\end{deluxetable*}

We obtained observations covering the NUV spectral range 
from the Barbara A.\ Mikulski Archive for Space Telescopes (MAST).
These observations were taken using the medium- or high-resolution
echelle gratings in the 
Space Telescope Imaging Spectrograph
(STIS; \citealt{kimble98,woodgate98})
on board the \textit{Hubble Space Telescope} (\textit{HST}).
Spectra downloaded from the MAST have been
reduced by the \textit{calstis} pipeline and
combined by \citet{ayers10}. 
Spectra obtained previously by our own observing programs
were reduced by the \textit{calstis} pipeline
and processed as described in
\citet{roederer12d,roederer14d} and \citet{placco14}.

\begin{figure*}
\begin{center}
\includegraphics[angle=270,width=6.0in]{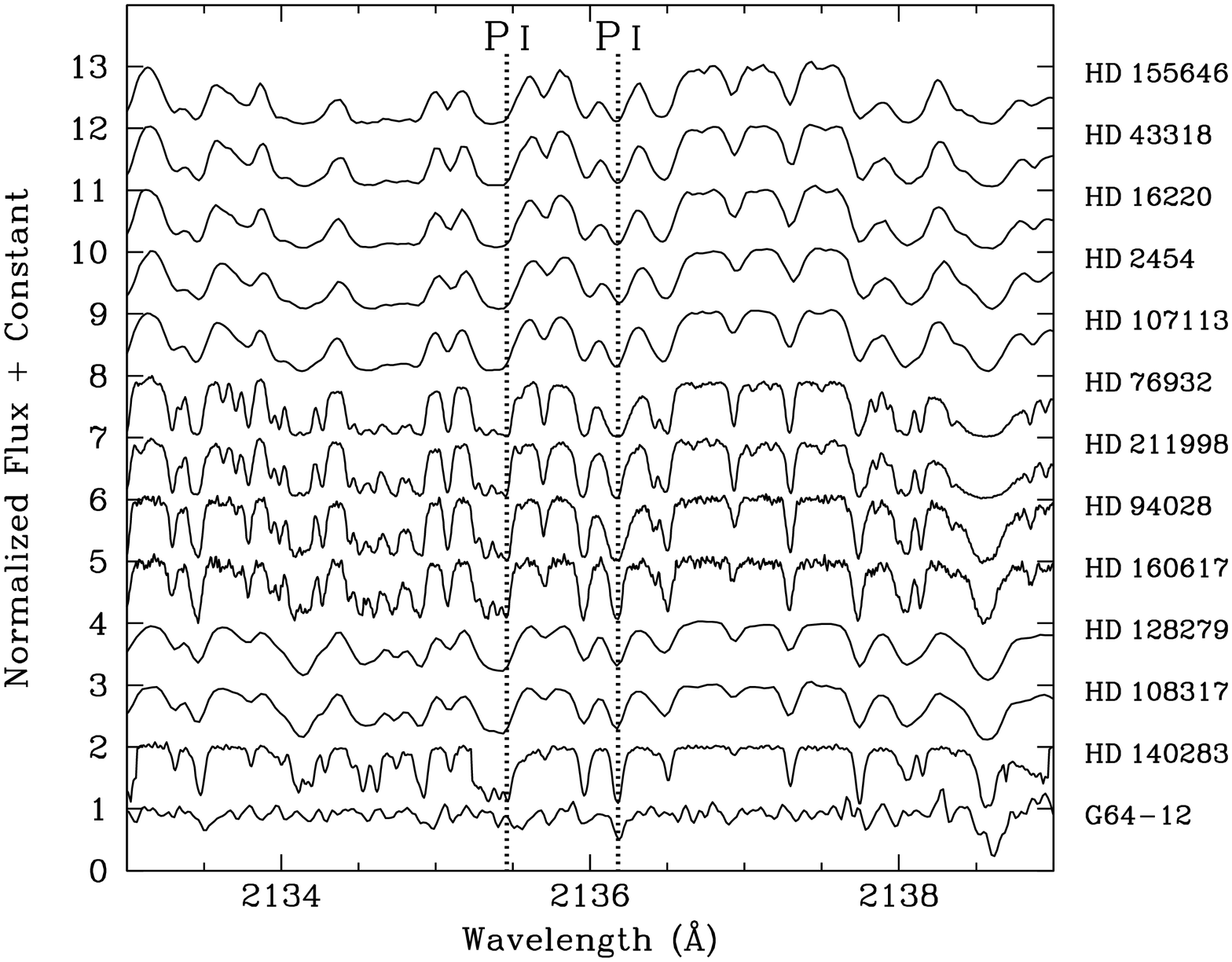}
\end{center}
\caption{
\label{specplot}
Spectral region surrounding the P~\textsc{i} 2135.46 and 2136.18~\AA\ lines.
The STIS spectra have been shifted vertically by adding 
a constant to the normalized flux values.
The stars are ordered by decreasing metallicity from top to bottom.
The relatively clean region of continuum from 2136.6--2136.8~\AA,
which we use to match our observed and synthetic spectra
(see Section~\ref{analysis}),
is apparent.
 }
\end{figure*}

Some stars were discarded from our analysis
because their spectra had low signal-to-noise (S/N) ratios.
Cooler or more metal-rich stars were discarded because we could
not reliably identify the continuum.
Table~\ref{obstab} presents a list of
the datasets and observational characteristics
for the 14~stars retained for our study.
Figure~\ref{specplot} illustrates the STIS spectra
of 13 of these stars for a region surrounding the
P~\textsc{i} 2135.46 and 2136.18~\AA\ lines.
No spectrum covering these lines exists for \bd,
so this star is not shown in Figure~\ref{specplot}.

\begin{deluxetable*}{cccccccc}
\tablecaption{Characteristics of Archival Optical Spectra
\label{vistab}}
\tablewidth{0pt}
\tabletypesize{\scriptsize}
\tablehead{
\colhead{Star} &
\colhead{$\lambda$} &
\colhead{$R$} &
\colhead{S/N} &
\colhead{S/N} &
\colhead{Instrument} &
\colhead{Program} &
\colhead{P.I.} \\
\colhead{} &
\colhead{(\AA)} &
\colhead{$\equiv \lambda/\Delta\lambda$} &
\colhead{4500~\AA} &
\colhead{6000~\AA} &
\colhead{} &
\colhead{ID} &
\colhead{} 
}
\startdata
G64-12    & 3420-9150  &  38,000 &   160 & 230   & MIKE  & \nodata       & Frebel            \\ 
HD~2454   & 3780-6910  & 115,000 &    93 & 180   & HARPS & 074.C-0364(A) & Robichon          \\ 
HD~16220  & 4350-6860  &  48,000 &   650 & 630   & HIRES & H5aH          & Stephens          \\ 
HD~43318  & 3100-10426 &  74,000 &   465 & 370   & UVES  & 266.D-5655(A) & \tablenotemark{a} \\ 
HD~76932  & 3100-6800\tablenotemark{b}
                       &  49,000 &\nodata& 315   & UVES  & 067.D-0439(A) & Primas            \\ 
          &            &         &       &       & UVES  & 076.B-0055(A) & Silva             \\ 
          &            &         &       &       & UVES  & 266.D-5655(A) & \tablenotemark{a} \\ 
HD~94028  & 3650-8000  &  30,000 &   430 & 250   & Tull  & \nodata       & Roederer          \\ 
HD~107113 & 4350-6860  &  48,000 &   250 & 670   & HIRES & H2aH          & Boesgaard         \\ 
HD~108317 & 3340-8000  &  38,000 &   570 & 390   & MIKE  & \nodata       & Roederer          \\ 
HD~128279 & 3340-8000  &  38,000 &   880 & 750   & MIKE  & \nodata       & Roederer          \\ 
HD~140283 & 3360-9400  &  38,000 &   390 & 260   & MIKE  & \nodata       & Frebel            \\ 
HD~155646 & 5650-8090  &  48,000 &\nodata& 650   & HIRES & H47aH         & Boesgaard         \\ 
HD~160617 & 3100-6800\tablenotemark{b}
                       &  51,000 &\nodata& 375 & UVES    & 065.L-0507(A) & Primas            \\ 
HD~211998 & 3100-6800\tablenotemark{b}
                       &  74,000 &\nodata& 580 & UVES    & 266.D-5655(A) & \tablenotemark{a}  
\enddata
\tablenotetext{a}{Observed for inclusion in ``A High-Resolution Spectroscopic
  Atlas of Stars across the Hertzsprung-Russell Diagram''.}
\tablenotetext{b}{Includes gaps in the spectrum}
\end{deluxetable*}

We supplement these NUV spectra with optical spectra
taken with ground-based telescopes and
found in online archives.
Table~\ref{vistab}
provides details about these optical spectra.
These data were taken using the 
High Accuracy Radial velocity Planet Searcher (HARPS; \citealt{mayor03}) 
on the ESO 3.6~m Telescope at La Silla, 
the
Ultraviolet and Visual Echelle Spectrograph (UVES; \citealt{dekker00}) 
on the 8.2~m Very Large Telescope (VLT UT2, Kueyen) 
at Cerro Paranal, or 
the High Resolution Echelle Spectrometer (HIRES; \citealt{vogt94})
on the 10~m Keck~I Telescope at Mauna Kea.
We obtained pipeline-reduced individual exposures from the
archives, normalized the continua, and shifted each exposure 
to laboratory-zero velocity.

Some stars were not present in these archives at the time of
our query, and we have collected new optical data instead.
Some observations were made using the 
Magellan Inamori Kyocera Echelle (MIKE)
spectrograph \citep{bernstein03} on the
6.5~m Landon Clay Telescope (Magellan~II)
at Las Campanas.
These data were obtained and reduced using procedures
described in detail in \citet{roederer14c}.
Other observations were obtained
using the Robert G.\ Tull Coud\'{e} Spectrograph
\citep{tull95} on the 2.7~m Harlan J.\ Smith Telescope 
at McDonald Observatory.
These data were reduced and combined using the REDUCE software
package \citep{piskunov02} and standard IRAF procedures.
Characteristics of these observations are also given
in Table~\ref{vistab}.

\section{Stellar parameters}
\label{atmpars}

Although most stars in this sample have stellar parameters
available in the literature, we carry out a homogeneous
analysis based on archival optical spectra.  
We determine
stellar parameters 
using a classical spectroscopic analysis 
starting with the measurement of
equivalent widths, $W_{\lambda}$.
These measurements are presented in Table~\ref{ewtab}.
We use the ATLAS9 model atmosphere grid \citep{castelli03} 
and a recent version of MOOG \citep{sneden73}
that includes the
contribution of Rayleigh scattering from H~\textsc{i} 
in the source function \citep{sobeck11}.

We use the line list
compiled in \citet{roederer10b}.
However, we found few Fe~\textsc{i} and Fe~\textsc{ii}
lines from this list were
measurable in the spectrum of 
\object[HD 155646]{HD~155646}.  
For this star, we use the
line list from \citet{ramirez13}.  
We verify that these two line
lists yield similar stellar parameters and element abundances in
the analysis of other metal-rich stars, and no systematic errors are
introduced.

\begin{deluxetable}{ccccccc}
\tablewidth{0pt}
\tabletypesize{\scriptsize}
\tablecaption{Line List and Equivalent Widths
\label{ewtab}}
\tablehead{
\colhead{Species} &
\colhead{$\lambda$} &
\colhead{E.P.} &
\colhead{\loggf} &
\colhead{$W_{\lambda}$} &
\colhead{$W_{\lambda}$} &
\colhead{\ldots} \\
\colhead{} &
\colhead{(\AA)} &
\colhead{(eV)} &
\colhead{} &
\colhead{G64-12} &
\colhead{HD~2454} &
\colhead{\ldots} }
\startdata
 Na~\textsc{i}  & 5682.63 &   2.10 & $-$0.71 &\nodata&  57.8 &\ldots \\
 Na~\textsc{i}  & 5688.20 &   2.10 & $-$0.45 &\nodata&  85.0 &\ldots \\
 Na~\textsc{i}  & 5889.95 &   0.00 & $+$0.11 &  35.8 & 303.7 &\ldots \\
 Na~\textsc{i}  & 5895.92 &   0.00 & $-$0.19 &  19.9 & 257.6 &\ldots \\
 \vdots         & \vdots  & \vdots & \vdots  &\vdots &\vdots &       
\enddata
\tablecomments{
All $W_{\lambda}$ are given in units of m\AA.
The complete version of Table~\ref{ewtab} can be found in the
online edition of the journal.
Only an abbreviated version is shown here to illustrate its
form and content.}
\end{deluxetable}

The derived model parameters are listed in Table~\ref{modeltab}.
Effective temperatures (\teff) and
microturbulent velocities (\vt) are determined by minimizing trends
of Fe~\textsc{i} lines with excitation potential (E.P.)
and reduced equivalent width [\logrw].
Surface gravities (\logg) are determined
by forcing the mean iron abundances derived from
Fe~\textsc{i} and Fe~\textsc{ii} lines to agree.  
Once excitation
and ionization balance are achieved, we apply the empirical
temperature calibration described in \citet{frebel13}.
Stars with [Fe/H]~$> -$2.5 fall outside the
metallicity range over which this calibration was tested.  
We note,
however, that studies of stars more metal-rich than this have found
similar discrepancies between photometric and spectroscopic
\teff\ values (e.g., \citealt{johnson02}),
indicating that a calibration such as this is appropriate.
Furthermore, at the relatively warm temperatures of these stars
($>$~5200 K), the correction is small and comparable to the uncertainty
($<$~150 K).
We have three stars in common with \citet{nissen07} and two in common
with \citet{asplund06}, who derived \teff\ by fitting the wings of the
H$\beta$ and H$\alpha$ lines, respectively.
The mean differences ($-$95~$\pm$~57~K and $-$72~K, respectively) 
are comparable to 
the systematic zeropoint uncertainties of these different methods,
suggesting that our results are not wildly in error.

\begin{deluxetable}{ccccc}
\tablewidth{0pt}
\tabletypesize{\scriptsize}
\tablecaption{Stellar Parameters and Metallicities
\label{modeltab}}
\tablehead{
\colhead{Star} &
\colhead{\teff} &
\colhead{\logg} &
\colhead{\vt} &
\colhead{[Fe/H]} \\
\colhead{} &
\colhead{(K)} &
\colhead{[cgs]} &
\colhead{(\kmsec)} &
\colhead{(dex)} 
}
\startdata
BD$+$44~493\tablenotemark{a} & 5430 & 3.40 & 1.30 & $-$3.83  \\
G64-12      & 6445 & 4.35 & 1.38 & $-$3.28  \\
HD~2454     & 6547 & 3.90 & 2.00 & $-$0.42  \\
HD~16220    & 6232 & 3.80 & 1.70 & $-$0.31  \\
HD~43318    & 6182 & 3.60 & 1.45 & $-$0.24  \\
HD~76932    & 5870 & 3.80 & 1.15 & $-$0.92  \\
HD~94028    & 5880 & 4.00 & 1.17 & $-$1.61  \\
HD~107113   & 6410 & 3.65 & 1.65 & $-$0.59  \\
HD~108317   & 5170 & 2.35 & 1.60 & $-$2.44  \\
HD~128279   & 5175 & 2.60 & 1.50 & $-$2.37  \\
HD~140283   & 5665 & 3.25 & 1.45 & $-$2.65  \\
HD~155646   & 6295 & 3.75 & 1.80 & $-$0.17  \\
HD~160617   & 5935 & 3.45 & 1.50 & $-$1.93  \\
HD~211998   & 5320 & 3.30 & 1.20 & $-$1.50  
\enddata
\tablenotetext{a}{Parameters adopted from \citet{ito13}}
\end{deluxetable}

\citet{roederer14c} derived stellar parameters and abundances
for a large sample of metal-poor stars.
That study derived \teff\ and \vt\ by
requiring no correlation between derived Fe~\textsc{i}
abundances and E.P.\ and \logrw,
as we have done.
That study found typical statistical uncertainties
of $\approx$~50~K in \teff\ and $\approx$~0.1~\kmsec\ in \vt\
for stars in similar evolutionary states
to those in our sample.
The uncertainties in \teff\ and \vt\ when eliminating
correlations with E.P.\ and \logrw\ are similar
in our work here.
We adopt these values as the statistical uncertainties
in \teff\ and \vt.
The dispersion among Fe~\textsc{i} and Fe~\textsc{ii} lines
suggests that $\approx$~0.3~dex is a fair estimate of the
statistical uncertainty in \logg.
We assume that the systematic uncertainty in \teff\ 
could be as large as the correction applied to convert
the spectroscopic \teff\ to the photometric ($V-K$) scale
(Equation~1 and Figure~2 of \citealt{frebel13}).
\citeauthor{roederer14c}\ made comparisons for the
derived model atmosphere parameters for large numbers
of stars (see Table~9 there).
The dispersions in \logg\ and \vt\ values
were found to be $\approx$~0.35~dex and $\approx$~0.4~\kmsec\
for stars in class ``SG,'' 
which have similar evolutionary states to the stars in our sample.
We adopt these values as the systematic uncertainties in
\logg\ and \vt.

We have not rederived stellar parameters or abundances for \bd,
and it is not listed in Table~\ref{vistab}.
This star has been studied extensively by \citet{ito09,ito13},
\citet{placco14}, and \citet{roederer14a}.
We do not detect P~\textsc{i}
in this star, so we simply adopt the \citet{ito13}
model parameters and abundances.
We apply the same 
$+$0.23~dex
offset to the phosphorus upper limit
to correct for differences in the optical and NUV abundance scales
(see Section~\ref{offset}).

\section{Neutral phosphorus lines detectable in the NUV}
\label{lines}

\begin{figure*}
\begin{center}
\includegraphics[angle=0,width=3.1in]{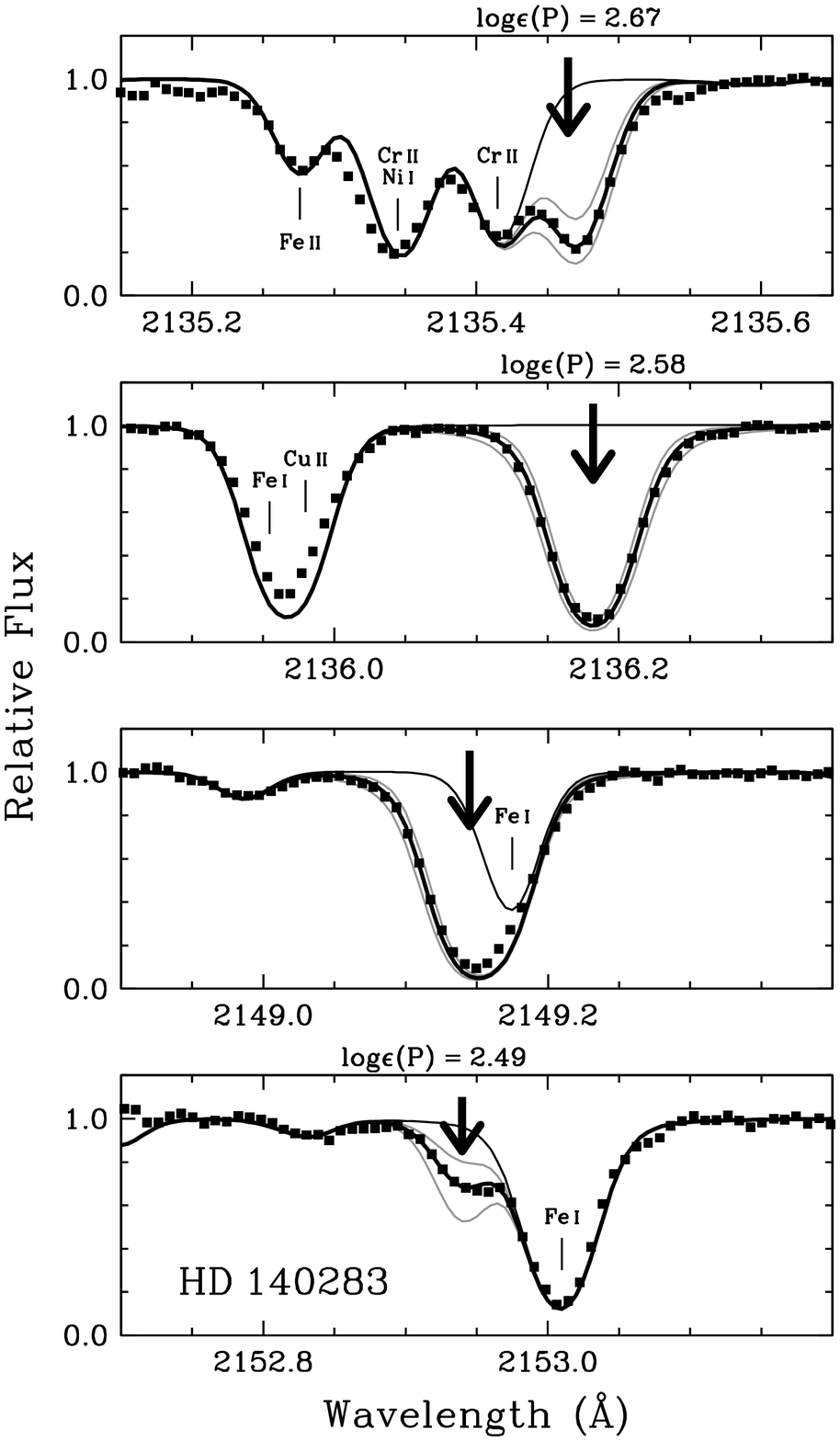}
\hspace*{0.1in}
\includegraphics[angle=0,width=3.1in]{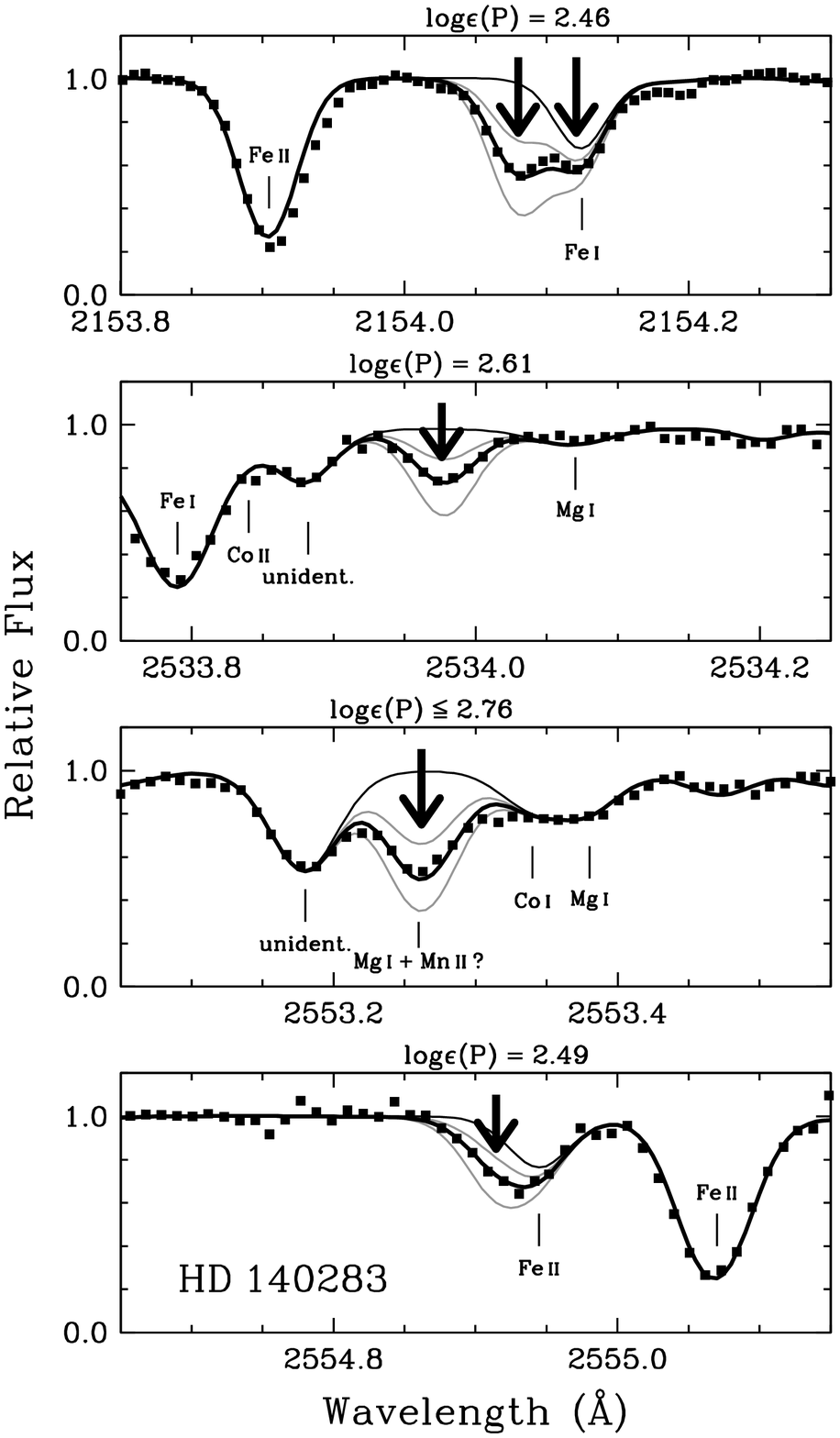}
\end{center}
\caption{
\label{synthplot}
Comparison of observed (points) and synthetic (curves)
spectra in \mbox{HD~140283}.
The bold black curve marks the best fit,
the thin gray curves mark variations in the best fit by $\pm$~0.3~dex,
and the thin black curve marks a synthesis
with no P~\textsc{i} absorption.
Arrows mark the position of P~\textsc{i} absorption lines.
Phosphorus abundances are indicated,
and the identities of blending features are marked.
 }
\end{figure*}

The high first-ionization potential of phosphorus
(10.49~eV) ensures that neutral phosphorus is the 
dominant ionization state in the atmospheres of late-type stars.
Several NUV P~\textsc{i} lines are 
listed in the Atomic Spectral Database (ASD) hosted by the
National Institute of Standards and Technology
(NIST; \citealt{kramida13}).
Among the stars in our sample,
\hdonefour\ has the most extensive spectral
coverage at $R \sim$~110,000 and high S/N.~
We use this star to identify which P~\textsc{i} lines
might be reliable abundance indicators.

The P~\textsc{i} 2135.465~\AA\ line is clearly detected
in the \hdonefour\ spectrum.
It is blended with a Cr~\textsc{ii} line at 2135.42~\AA.
This line is marginally useful as an abundance indicator
in \hdonefour.
It is illustrated in Figure~\ref{synthplot},
which compares the 
observed spectrum of \hdonefour\ 
with our synthetic spectra.
The abundances derived from this line
and all others in \hdonefour\
are presented in Table~\ref{atomictab}.

\begin{deluxetable*}{cccccccc}
\tablecaption{Atomic Data and Abundances
for P~\textsc{i} Lines Detected in HD~140283
\label{atomictab}}
\tablewidth{0pt}
\tabletypesize{\scriptsize}
\tablehead{
\colhead{Wavelength} &
\colhead{Upper} &
\colhead{$E_{\rm upper}$\tablenotemark{a}} &
\colhead{Lower} &
\colhead{$E_{\rm lower}$\tablenotemark{a}} &
\colhead{\loggf} &
\colhead{$\log\Gamma_{6}/N_{\rm H}$\tablenotemark{b}} &
\colhead{$\log \epsilon$} \\
\colhead{(\AA)} &
\colhead{level} &
\colhead{(cm$^{-1}$)} &
\colhead{level} &
\colhead{(cm$^{-1}$)} &
\colhead{} &
\colhead{(rad cm$^{3}$ s$^{-1}$)} &
\colhead{HD~140283\tablenotemark{c}} }
\startdata
 2135.465 & $(^{3}P)4s ^{2}P_{3/2}$ & 58174.366 & $3p^{3}~^{2}D^{o}_{3/2}$ & 11361.02 & $-$1.24 & $-$7.42 & 2.67    \\
 2136.182 & $(^{3}P)4s ^{2}P_{3/2}$ & 58174.366 & $3p^{3}~^{2}D^{o}_{5/2}$ & 11376.63 & $-$0.11 & $-$7.42 & 2.58    \\
 2149.145 & $(^{3}P)4s ^{2}P_{1/2}$ & 57876.574 & $3p^{3}~^{2}D^{o}_{3/2}$ & 11361.02 & $-$0.36 & $-$7.42 & \nodata \\
 2152.940 & $(^{1}D)4s ^{2}D_{3/2}$ & 65156.242 & $3p^{3}~^{2}P^{o}_{1/2}$ & 18722.71 & $-$0.87 & $-$7.22 & 2.49    \\
 2154.080 & $(^{1}D)4s ^{2}D_{5/2}$ & 65157.126 & $3p^{3}~^{2}P^{o}_{3/2}$ & 18748.01 & $-$0.62 & $-$7.22 & 2.46    \\
 2154.121 & $(^{1}D)4s ^{2}D_{3/2}$ & 65156.242 & $3p^{3}~^{2}P^{o}_{3/2}$ & 18748.01 & $-$1.32 & $-$7.22 & \tablenotemark{d} \\
 2533.976 & $(^{3}P)4s ^{2}P_{3/2}$ & 58174.366 & $3p^{3}~^{2}P^{o}_{1/2}$ & 18722.71 & $-$1.11 & $-$7.42 & 2.61    \\
 2535.603 & $(^{3}P)4s ^{2}P_{3/2}$ & 58174.366 & $3p^{3}~^{2}P^{o}_{3/2}$ & 18748.01 & $-$0.44 & $-$7.42 & \nodata \\
 2553.262 & $(^{3}P)4s ^{2}P_{1/2}$ & 57876.574 & $3p^{3}~^{2}P^{o}_{1/2}$ & 18722.71 & $-$0.86 & $-$7.42 & $\leq$~2.76 \\
 2554.915 & $(^{3}P)4s ^{2}P_{1/2}$ & 57876.574 & $3p^{3}~^{2}P^{o}_{3/2}$ & 18748.01 & $-$1.23 & $-$7.42 & 2.49    
\enddata
\tablenotetext{a}{\citet{svendenius80}}
\tablenotetext{b}{Calculated for $T=$~10$^{4}$~K}
\tablenotetext{c}{Corrected for optical-NUV offset; see Section~\ref{offset}}
\tablenotetext{d}{{Line} forms a blended, single feature with 
P~\textsc{i} 2154.08~\AA\ line}
\end{deluxetable*}

The P~\textsc{i} 2136.182~\AA\ line is 
approaching saturation in \hdonefour.
We see no evidence of line asymmetries 
or hints of absorption by other species
in the observed line profile of \hdonefour.
Saturated lines are less sensitive
to the abundance,
but otherwise this line appears to be a
reliable abundance indicator.
Figure~\ref{synthplot} illustrates this line in \hdonefour.

The P~\textsc{i} 2149.145~\AA\ line is detected but blended
with an Fe~\textsc{i} line at 2149.18~\AA.
This composite feature is saturated in \hdonefour\ 
and is not useful as a phosphorus abundance indicator.
This line is shown in Figure~\ref{synthplot}, 
but the synthetic spectrum that has been
``fit'' to the observed one
has not been used to derive an abundance.

The P~\textsc{i} 2152.940~\AA\ line 
is located in the wing of a much stronger
Fe~\textsc{i} line at 2153.01~\AA.
It can be used as an abundance indicator
in \hdonefour,
as illustrated in Figure~\ref{synthplot}.

The P~\textsc{i} 2154.080 and 2154.121~\AA\ 
transitions blend together to form one line,
although the absorption is dominated by the 2154.080~\AA\ line.
They are blended with an Fe~\textsc{i} line at 2154.13~\AA\
of comparable strength.
NIST reports a ``C'' accuracy (25\%) for the 
transition probability of this Fe~\textsc{i} line.
We adjust its \loggf\ value in our synthesis 
(to $-$2.57)
to derive an 
approximate phosphorus abundance in \hdonefour.
Figure~\ref{synthplot} shows our best
attempt to fit this blended feature in \hdonefour.

The P~\textsc{i} 2533.976~\AA\ line is clean, unsaturated, and
easily detected in \hdonefour.
An unidentified absorption feature is present
at 2533.89~\AA.
We model this feature (and all other unidentified
absorption lines) as an Fe~\textsc{i} line
with an E.P.\ of 1.5~eV and \loggf\ value constrained 
by the observed line profile.
A weak Mg~\textsc{i} absorption feature
may be present at 2534.07~\AA.
The P~\textsc{i} 2533.976~\AA\ line
appears to be a reliable abundance indicator in 
\hdonefour, as shown in Figure~\ref{synthplot}.
Tests conducted using $R \sim$~30,000 spectra of
\hdoneoh\ and \hdonetwo\
(obtained in program GO-12268)
indicate, unfortunately, that the P~\textsc{i} 2533.976~\AA\
line is too blended to serve as a reliable abundance indicator
in lower-resolution data.
Both \hdoneoh\ and \hdonetwo\ have
[Fe/H]~$< -$2.4, so the P~\textsc{i} 2533.976~\AA\ line
will not be a reliable abundance indicator in
higher-metallicity stars, either.

The P~\textsc{i} 2535.603~\AA\ line cannot be discerned amid
the absorption from a much stronger Fe~\textsc{i} line
at 2535.61~\AA.
Our synthesis suggests that this line
would be detectable in the absence of the Fe~\textsc{i} line.
We cannot use this P~\textsc{i} line as an abundance indicator,
and it is not illustrated in Figure~\ref{synthplot}.

The P~\textsc{i} 2553.262~\AA\ line is detected but blended with
several features of comparable strength.
An unidentified feature at 2553.18~\AA\ and a
Co~\textsc{i} feature at 2553.34~\AA\ can be
fit to match the observed line profile.
Our synthesis also suggests that absorption from
Mg~\textsc{i} and Mn~\textsc{ii} could be 
present at 2553.26~\AA, 
and both features are coincident with the 
P~\textsc{i} line.
The \loggf\ values of these blending features
are not known from laboratory measurements.
A synthesis with both the Mg~\textsc{i} and Mn~\textsc{ii}
transitions removed can generally reproduce
the P~\textsc{i} line profile.
Since the Mg~\textsc{i} and Mn~\textsc{ii} transitions
could account for an unknown fraction of the absorption at this wavelength,
we derive only
an upper limit on the
phosphorus abundance.
This upper limit is larger than the phosphorus
abundances derived from other, more secure, abundance indicators.
Figure~\ref{synthplot} illustrates this fit,
which is computed assuming that
neither Mg~\textsc{i} nor Mn~\textsc{ii}
contribute to the absorption.

Absorption is detected at the P~\textsc{i} 2554.915~\AA\ line.
A neighboring line of Fe~\textsc{ii} at 2554.94~\AA\ has 
comparable strength, and it blends with
the P~\textsc{i} line.
NIST reports a ``D'' accuracy (50\%) for the
transition probability of this Fe~\textsc{ii} line.
This P~\textsc{i} line 
is marginally useful as an abundance indicator
in \hdonefour\
and is illustrated in Figure~\ref{synthplot}.

In summary, there are six marginally useful P~\textsc{i} 
abundance indicators in the NUV spectrum of \hdonefour.
Our final phosphorus abundance in \hdonefour\ is a 
mean of these six abundance derivations.
In stars that are more metal-rich or cooler than
\hdonefour, or in spectra taken at lower 
resolution or with lower S/N levels,
most of these lines are not useful as abundance indicators.
The phosphorus abundances we derive for the other stars
in our sample rely only on the P~\textsc{i} 2136~\AA\ line.
No existing NUV spectrum of \bd, however, 
covers the P~\textsc{i} 2136~\AA\ line.
We derive an upper limit on the phosphorus
abundance in \bd\ from the P~\textsc{i} 2553~\AA\ line
assuming that no Mg~\textsc{i} or Mn~\textsc{ii}
contribute to the absorption.

\section{Atomic data for phosphorus}
\label{atomic}

\subsection{Transition probabilities}
\label{loggfvalues}

Seven of the 10 P~\textsc{i} transitions we examine in \hdonefour\
connect to levels in the $^{2}P$ multiplet,
and three connect to levels in the
$^{2}D$ multiplet.
The upper and lower levels of each transition
are listed in Table~\ref{atomictab}.
Relatively little experimental work has been 
done on these P~\textsc{i} transitions.
\citet{savage66} measured the radiative lifetime, $\tau$, 
of the upper $^{2}P$ multiplet by the phase-shift method,
finding $\tau =$~2.4~$\pm$~0.3~ns.
This agrees well with theoretical calculations by
\citet{lawrence67},
$\tau =$~2.38~ns.
\citeauthor{savage66} also measured the lifetime 
of the $^{2}D$ multiplet,
2.9~$\pm$~0.4~ns,
which agrees with that measured using the beam-foil method
by \citet{curtis71}, $\tau =$~3.6~$\pm$~0.4~ns.
The \citeauthor{lawrence67} theoretical calculation
gives $\tau =$~2.89~ns for this level.
No experimental measures of the 
branching fractions (BF) are reported in the literature 
for any of the transitions of interest.

NIST reports \loggf\ values for these transitions 
that normalize the \citet{lawrence67} BF calculations
to the \citet{savage66} lifetimes.
NIST estimates a ``C'' accuracy for these values (25\% or 
$\approx$~0.12~dex).
We infer the BFs from \citeauthor{lawrence67}'s calculations
using the relation $A_{ki} = {\rm BF}/\tau$,
where $A_{ki}$ is the transition probability.
The P~\textsc{i} 2136~\AA\ line, 
which we use to derive the phosphorus abundance
in all stars in our study,
is a dominant branch (68\%)
from the $^{2}P_{3/2}$ level.
Its \loggf\ value is likely to be the
least uncertain of the transitions considered.
The other transitions
have BFs ranging from 5\% to 17\%.

Lacking additional information,
we simply adopt the \loggf\ values
reported by NIST for all P~\textsc{i} transitions
in our study.
These values are listed in Table~\ref{atomictab}.

\subsection{The damping constant for the 2136~\AA\ transition}
\label{damping}

The P~\textsc{i} 2136~\AA\ line shows
damping wings in most stars in our sample.
The strength of a spectral line with damping wings is
sensitive to the number of absorbers,
the oscillator strength, and the sum of the damping constants
from all sources of broadening.
Collisions with neutral hydrogen 
dominate the damping in late-type stars,
so the van der Waals broadening constant, 
$\Gamma_{6}$, must be estimated.
We adopt damping constants from the
calculations of \citet{barklem00} and \citet{barklem05}
when such values are available; however,
these studies have not reported
damping constants for any of 
the P~\textsc{i} lines used in our analysis.
We calculate damping constants 
using the same theory 
\citep{anstee95}
for all 10~P~\textsc{i} lines
examined in our study.
These values are presented in Table~\ref{atomictab}.

When compared
with the damping constants predicted by the 
\citet{unsold55} approximation,
the values in Table~\ref{atomictab} are 
larger by factors of 1.97 (for the 2136~\AA\ line)
to 2.40.
These enhancements are in line with
empirical tests using Fe~\textsc{i} lines
in the Sun (e.g., \citealt{cowley69,simmons82,ryan98}).
Comparisons of the \citet{barklem00}
calculations with the \citeauthor{unsold55}
approximation suggest similar enhancements,
typically factors of 1 to 2
for lines with E.P.\
between 1 and 2~eV.

To estimate the uncertainty in the
damping constant for the P~\textsc{i} 2136~\AA\ line,
we independently constrain the damping constant
using abundances derived from other
P~\textsc{i} lines whose abundances are insensitive
to the damping wings in \hdonefour.
If the oscillator strength and abundance are known,
the damping constant of a given line
can be inferred from its
observed equivalent width or line profile.
The mean phosphorus abundance 
derived from the other P~\textsc{i} lines
is lower than that derived from the 2136~\AA\ line by 
0.19~dex if no enhancement 
to the \citet{unsold55}
approximation is adopted.
Enhancement by a factor of 2 reduces 
the difference between the abundance derived from
the 2136~\AA\ line and the other lines to 0.14~dex,
which is within the 1$\sigma$
abundance dispersion found for the other
P~\textsc{i} lines in \hdonefour, 0.16~dex.
For small enhancement factors, 
each factor of 2 enhancement to the \citet{unsold55}
approximation in \hdonefour\
reduces the derived abundance from the P~\textsc{i} 2136~\AA\
line by 0.05~dex.
We conclude that the uncertainty in the 
damping constant for the P~\textsc{i} 2136~\AA\ line 
contributes less than 0.05~dex
to the total abundance uncertainty.

\section{Abundance analysis}
\label{analysis}

We employ standard methods to derive abundances of phosphorus
and other elements of interest.
Abundances of iron, sodium, magnesium, 
aluminum, silicon, and calcium
are derived using equivalent widths
(measured from only the optical spectra),
model atmospheres interpolated from the ATLAS9 grid,
and the line analysis code MOOG.~
These results are presented in Table~\ref{elemtab}.
The equivalent width,
wavelength, E.P.,
and \loggf\ value of each line 
are presented in Table~\ref{ewtab}.
All damping constants are taken from
\citet{barklem00} and \citet{barklem05},
when available,
otherwise we resort to the \citet{unsold55} approximation
for these lines.

\begin{deluxetable*}{ccccccccccccc}
\tablewidth{0pt}
\tabletypesize{\tiny}
\tablecaption{Optical Abundances of Fe, Na, Mg, Al, Si, and Ca
\label{elemtab}}
\tablehead{
\colhead{Star} &
\colhead{[Fe/H]} & 
\colhead{$N$} &
\colhead{[Na/Fe]} &
\colhead{$N$} &
\colhead{[Mg/Fe]} &
\colhead{$N$} &
\colhead{[Al/Fe]} &
\colhead{$N$} &
\colhead{[Si/Fe]} &
\colhead{$N$} &
\colhead{[Ca/Fe]} &
\colhead{$N$} 
}
\startdata
BD$+$44~493\tablenotemark{a}
            & $-$3.83 (0.19) &\ldots&$+$0.30 (0.10)&\ldots&$+$0.46 (0.08)&\ldots&$-$0.56 (0.14)&\ldots&$+$0.49 (0.14)&\ldots&$+$0.31 (0.10) &\ldots\\
G64-12      & $-$3.28 (0.09) &  57 & $-$0.05 (0.13) &  2 & $+$0.37 (0.11) &  3 & $-$0.58 (0.18) &  1 & $+$0.19 (0.22) &  1 & $+$0.48 (0.14) &  8 \\
HD~2454     & $-$0.42 (0.12) & 172 & $+$0.05 (0.15) &  4 & $+$0.07 (0.15) &  3 & $+$0.15 (0.21) &  1 & $+$0.24 (0.22) &  4 & $+$0.07 (0.16) & 19 \\
HD~16220    & $-$0.31 (0.11) & 121 & $+$0.21 (0.15) &  2 & $+$0.12 (0.16) &  1 & \nodata        &  0 & $+$0.04 (0.22) &  2 & $+$0.19 (0.16) & 15 \\
HD~43318    & $-$0.24 (0.11) & 189 & $+$0.24 (0.15) &  2 & $-$0.01 (0.13) &  3 & $-$0.23 (0.18) &  2 & $-$0.12 (0.22) &  3 & $+$0.18 (0.16) & 21 \\
HD~76932    & $-$0.92 (0.12) & 140 & $+$0.14 (0.16) &  2 & $+$0.24 (0.15) &  2 & \nodata        &  0 & $+$0.05 (0.22) &  3 & $+$0.32 (0.16) & 18 \\
HD~94028    & $-$1.61 (0.13) & 234 & $+$0.25 (0.17) &  3 & $+$0.31 (0.15) &  7 & $-$0.80 (0.22) &  1 & $+$0.27 (0.24) &  2 & $+$0.40 (0.17) & 26 \\
HD~107113   & $-$0.59 (0.11) & 135 & $+$0.31 (0.15) &  2 & $+$0.20 (0.16) &  1 & \nodata        &  0 & $+$0.03 (0.22) &  2 & $+$0.22 (0.16) & 20 \\
HD~108317   & $-$2.44 (0.11) & 243 & $+$0.62 (0.15) &  2 & $+$0.44 (0.13) &  7 & $-$1.10 (0.20) &  1 & $+$0.59 (0.23) &  1 & $+$0.44 (0.16) & 26 \\
HD~128279   & $-$2.37 (0.12) & 253 & $+$0.26 (0.15) &  4 & $+$0.41 (0.14) &  7 & $-$0.93 (0.21) &  1 & $+$0.46 (0.22) &  3 & $+$0.42 (0.16) & 26 \\
HD~140283   & $-$2.65 (0.08) & 165 & $+$0.24 (0.12) &  2 & $+$0.30 (0.10) &  6 & $-$0.89 (0.17) &  1 & $+$0.42 (0.21) &  1 & $+$0.33 (0.14) & 20 \\
HD~155646   & $-$0.16 (0.07) &  64 & $+$0.20 (0.11) &  2 & $-$0.01 (0.11) &  1 & $-$0.21 (0.15) &  2 & $+$0.21 (0.19) &  2 & $+$0.10 (0.13) &  8 \\
HD~160617   & $-$1.92 (0.08) & 133 & $+$0.53 (0.11) &  4 & $+$0.27 (0.10) &  4 & \nodata        &  0 & $+$0.37 (0.21) &  1 & $+$0.42 (0.14) & 19 \\
HD~211998   & $-$1.50 (0.11) & 123 & $-$0.11 (0.15) &  2 & $+$0.30 (0.14) &  2 & \nodata        &  0 & $+$0.14 (0.22) &  3 & $+$0.39 (0.16) & 17 
\enddata
\tablenotetext{a}{Abundances adopted from \citet{ito13}}
\end{deluxetable*}

All phosphorus abundances are derived by spectrum synthesis matching,
and we use MOOG to generate the synthetic spectra.
The best fits are determined by 
minimizing the 
residuals between the observed and synthetic spectra.
There is a relatively clean region of continuum 
from 2136.6--2136.8~\AA\ in most of our spectra.
Our fits match the observed and synthetic spectra levels here.
Only in the most metal-rich stars
does this region begin to show weak absorption,
but our synthetic spectra adequately reproduce this region.
The derived phosphorus abundances are reported in Table~\ref{phostab}.

\begin{deluxetable}{ccc}
\tablewidth{0pt}
\tabletypesize{\scriptsize}
\tablecaption{Phosphorus Abundances
\label{phostab}}
\tablehead{
\colhead{Star} &
\colhead{$\log\epsilon$ (P)} &
\colhead{[P/Fe]} 
}
\startdata
BD$+$44~493 & $<$~1.83        & $<+$0.25        \\
G64-12      &     2.28 (0.34) &  $+$0.15 (0.26) \\
HD~2454     &     5.06 (0.27) &  $+$0.07 (0.21) \\
HD~16220    &     4.91 (0.26) &  $-$0.19 (0.20) \\
HD~43318    &     5.03 (0.26) &  $-$0.14 (0.20) \\
HD~76932    &     4.93 (0.29) &  $+$0.44 (0.20) \\
HD~94028    &     4.41 (0.29) &  $+$0.61 (0.21) \\
HD~107113   &     5.06 (0.26) &  $+$0.24 (0.20) \\
HD~108317   &     2.83 (0.35) &  $-$0.14 (0.20) \\
HD~128279   &     3.08 (0.36) &  $+$0.04 (0.22) \\
HD~140283   &     2.55 (0.30) &  $-$0.21 (0.19) \\
HD~155646   &     5.11 (0.25) &  $-$0.14 (0.19) \\
HD~160617   &     3.56 (0.27) &  $+$0.07 (0.17) \\
HD~211998   &     3.78 (0.33) &  $-$0.13 (0.21) 
\enddata
\end{deluxetable}

We assess the sensitivity of the phosphorus abundances to 
uncertainties in the model atmosphere parameters
by altering each parameter by a fixed amount.
Three sets of numbers are presented in Table~\ref{modelsigmatab}.
One set corresponds to the metal-rich stars
observed with moderately-high spectral resolution 
($R \sim$~30,000),
one set corresponds to the metal-poor stars
observed with moderately-high spectral resolution
($R \sim$~30,000), and
one set corresponds to the metal-poor stars
observed with high spectral resolution
($R \sim$~110,000).
The stars in each group are identified in the footnotes
to Table~\ref{modelsigmatab}.
The uncertainties reported in Table~\ref{modelsigmatab}
are conservative since covariances among the
model parameters are not taken into account
and could reduce the abundance uncertainties by small amounts.
Regardless, the abundance uncertainties are not
very large ($\lesssim$~0.1~dex),
and they do not dominate the error budget.

\begin{deluxetable}{ccccc}
\tablewidth{0pt}
\tabletypesize{\scriptsize}
\tablecaption{Abundance Sensitivities
to Model Atmosphere Uncertainties
\label{modelsigmatab}}
\tablehead{
\colhead{Group} &
\colhead{Abundance} &
\colhead{$\Delta\teff =$} &
\colhead{$\Delta\logg =$} &
\colhead{$\Delta\vt =$~} \\
\colhead{} &
\colhead{or Ratio} &
\colhead{$\pm$~100~K} &
\colhead{$\pm$~0.3~dex} &
\colhead{$\pm$~0.2~\kmsec}
}
\startdata
Metal-rich,                       & [Na/Fe] & $\mp$~0.02 & $\pm$~0.03 & $\mp$~0.03 \\
$R \sim$~30,000\tablenotemark{a}  & [Mg/Fe] & $\mp$~0.01 & $\pm$~0.01 & $\mp$~0.02 \\
                                  & [Al/Fe] & $\mp$~0.04 & $\pm$~0.01 & $\mp$~0.05 \\
                                  & [Si/Fe] & $\mp$~0.04 & $\mp$~0.02 & $\mp$~0.04 \\
                                  & [P/Fe]  & $\mp$~0.01 & $\mp$~0.02 & $\mp$~0.02 \\
                                  & [Ca/Fe] & $\mp$~0.01 & $\pm$~0.03 & $\mp$~0.01 \\
                                  & $\log\epsilon$(P)  & $\pm$~0.07 & $\mp$~0.03 & $\mp$~0.04 \\
                                  & $\log\epsilon$(Fe) & $\pm$~0.08 & $\mp$~0.03 & $\mp$~0.06 \\
\hline
Metal-poor,                       & [Na/Fe] & $\mp$~0.01 & $\pm$~0.04 & $\mp$~0.01 \\
$R \sim$~30,000\tablenotemark{b}  & [Mg/Fe] & $\mp$~0.02 & $\pm$~0.02 & $\mp$~0.01 \\
                                  & [Al/Fe] & $\mp$~0.01 & $\pm$~0.03 & $\pm$~0.02 \\
                                  & [Si/Fe] & $\mp$~0.05 & $\mp$~0.02 & $\mp$~0.03 \\
                                  & [P/Fe]  & $\mp$~0.05 & $\pm$~0.02 & $\pm$~0.03 \\
                                  & [Ca/Fe] & $\mp$~0.04 & $\pm$~0.01 & $\mp$~0.02 \\
                                  & $\log\epsilon$(P)  & $\pm$~0.06 & $\mp$~0.05 & $\mp$~0.08 \\
                                  & $\log\epsilon$(Fe) & $\pm$~0.11 & $\mp$~0.03 & $\mp$~0.05 \\
\hline
Metal-poor,                       & [Na/Fe] & $\mp$~0.02 & $\pm$~0.01 & $\mp$~0.02 \\
$R \sim$~110,000\tablenotemark{c} & [Mg/Fe] & $\mp$~0.02 & $\pm$~0.03 & $\mp$~0.03 \\
                                  & [Al/Fe] & $\pm$~0.01 & $\pm$~0.05 & $\mp$~0.01 \\
                                  & [Si/Fe] & $\mp$~0.04 & $\pm$~0.01 & $\mp$~0.03 \\
                                  & [P/Fe]  & $\mp$~0.03 & $\pm$~0.01 & $\mp$~0.01 \\
                                  & [Ca/Fe] & $\mp$~0.03 & $\pm$~0.01 & $\mp$~0.01 \\
                                  & $\log\epsilon$(P)  & $\pm$~0.07 & $\mp$~0.05 & $\mp$~0.05 \\
                                  & $\log\epsilon$(Fe) & $\pm$~0.10 & $\mp$~0.04 & $\mp$~0.05 
\enddata
\tablenotetext{a}{\mbox{HD~2454}, \mbox{HD~16220},
\mbox{HD~43318}, \mbox{HD~107113}, \mbox{HD~155646}}
\tablenotetext{b}{\mbox{G64-12}, \mbox{HD~108317},
\mbox{HD~128279}}
\tablenotetext{c}{\mbox{HD~76932}, \mbox{HD~94028}, 
\mbox{HD~140283}, \mbox{HD~160617}, \mbox{HD~211998}}
\end{deluxetable}

We have also
considered the effect of macroturbulence 
in the line profiles and derived abundances.
We adopt macroturbulent velocities
from Figure~3 of \citet{valenti05}
and compute the macroturbulence profiles 
using the radial-tangential formalism 
presented by \citet{gray92}.
These provide a reasonable fit to the line profiles
observed in our high-resolution NUV spectra.
Gaussian smoothing of our synthetic spectra
also provides a reasonable fit.
The derived abundances are not sensitive to these values.

\subsection{The offset between the optical and NUV abundance scales}
\label{offset}

Previous studies of abundances derived from NUV spectra of late-type
stars have uncovered evidence that the optical and NUV 
abundance scales may differ by small amounts.
For example, the [Fe/H] ratio derived from Fe~\textsc{i} lines
in the optical is frequently higher than the [Fe/H] ratio
derived from Fe~\textsc{i} lines in the NUV.~
Furthermore, the effect is most pronounced at wavelengths in the
blue near the transition from the Paschen continuum
to the Balmer continuum.
\citet{roederer10a,roederer12d,roederer14d}, \citet{lawler13}, 
\citet{wood13,wood14}, and \citet{placco14} have examined
and discussed this issue.
Presently, the origin of the discrepancy is unknown,
although missing sources of continuous opacity
and departures from local
thermodynamic equilibrium (LTE) 
may each be responsible to some degree.

We have derived abundances from 
5--17
Fe~\textsc{i} and \textsc{ii} lines 
in the NUV in six stars to get a sense of the magnitude
of this offset in our data.
These data are compared with the mean abundance derived from 
Fe~\textsc{i} and \textsc{ii} lines in the optical.
Table~\ref{offsettab} presents these results.
These six stars are selected to represent the
range of metallicities, stellar parameters,
and data quality of our full sample.
We did not derive abundances from
sufficient numbers of Fe~\textsc{i} and \textsc{ii}
lines in the NUV in the other stars
to warrant inclusion in this sample.

On average, the 
Fe~\textsc{i} lines in the optical
yield abundances higher by 
$+$0.30~$\pm$~0.02~dex ($\sigma =$~0.05~dex).
The Fe~\textsc{ii} lines in the optical yield abundances higher by
$+$0.23~$\pm$~0.04~dex ($\sigma =$~0.09~dex).
The difference between these two values, $+$0.07~dex,
suggests that non-LTE effects could be preferentially
affecting low-lying levels of the minority species, Fe~\textsc{i},
which are commonly found in the NUV.~
We observe an offset between the optical and NUV
abundances derived from Fe~\textsc{ii} lines, too.
The Fe~\textsc{ii} electronic level populations should not
deviate substantially from their LTE values.
This suggests that non-LTE effects are probably not the
dominant source of the offset between the optical and NUV
abundance scales.

\begin{deluxetable*}{cccccccccccc}
\tablewidth{0pt}
\tabletypesize{\tiny}
\tablecaption{Mean Fe~\textsc{i} Abundances Derived from Optical and NUV Lines
\label{offsettab}}
\tablehead{
\colhead{Star} &
\colhead{$\langle A$(Fe~\textsc{i})$\rangle$} &
\colhead{$N$} &
\colhead{$\langle A$(Fe~\textsc{i})$\rangle$} &
\colhead{$N$} &
\colhead{$\Delta$(Fe~\textsc{i})} &
\colhead{$\langle A$(Fe~\textsc{ii})$\rangle$} &
\colhead{$N$} &
\colhead{$\langle A$(Fe~\textsc{ii})$\rangle$} &
\colhead{$N$} &
\colhead{$\Delta$(Fe~\textsc{ii})} &
\colhead{$\Delta$(Fe~\textsc{i}$-$Fe~\textsc{ii})} \\
\colhead{} &
\colhead{opt} &
\colhead{} &
\colhead{NUV} &
\colhead{} &
\colhead{opt$-$NUV} &
\colhead{opt} &
\colhead{} &
\colhead{NUV} &
\colhead{} &
\colhead{opt$-$NUV} &
\colhead{}
}
\startdata
 HD~2454   & 7.08 & 172 & 6.69 & 16 & $+$0.39 & 7.09 & 18 & 6.75 & 12 & $+$0.34 & $+$0.05 \\
 HD~16220  & 7.19 & 121 & 6.94 & 11 & $+$0.25 & 7.20 & 13 & 6.88 &  7 & $+$0.32 & $-$0.07 \\
 HD~43318  & 7.26 & 189 & 6.99 & 13 & $+$0.27 & 7.26 & 20 & 7.05 &  7 & $+$0.21 & $+$0.06 \\
 HD~108317 & 5.06 & 243 & 4.73 & 17 & $+$0.33 & 5.07 & 25 & 4.81 & 13 & $+$0.26 & $+$0.07 \\
 HD~140283 & 4.85 & 170 & 4.58 &  5 & $+$0.27 & 4.87 & 15 & 4.78 &  5 & $+$0.09 & $+$0.18 \\
 HD~155646 & 7.34 &  65 & 7.02 & 11 & $+$0.32 & 7.35 &  5 & 7.15 &  7 & $+$0.20 & $+$0.12 \\
\hline
   & & & & $\langle\Delta\rangle =$ & $+$0.30 &&&& $\langle\Delta\rangle =$& $+$0.23 & $+$0.07 \\
   & & & & s.e.m.~$=$               &    0.02 &&&& s.e.m.~$=$              &    0.03 &    0.03 
\enddata
\end{deluxetable*}

\begin{figure}
\begin{center}
\includegraphics[angle=0,width=3.1in]{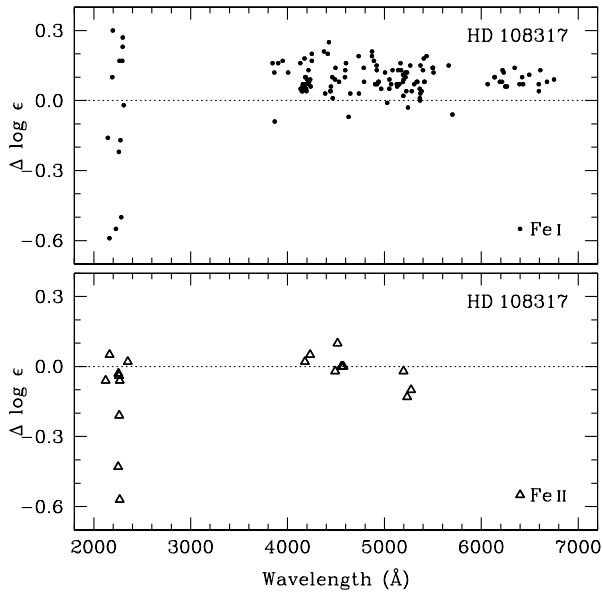}
\end{center}
\caption{
\label{fewaveplot}
Differences in derived line-by-line abundances
in \mbox{HD~108317} between \citet{roederer12d,roederer14d}
and our study.
Differences are in the sense of our study minus
\citeauthor{roederer14d} 
The top panel shows Fe~\textsc{i} lines, and the bottom panel
shows Fe~\textsc{ii} lines.
The dotted line marks a difference of zero.
 }
\end{figure}

\citet{roederer12d,roederer14d} also performed a detailed analysis of the
Fe~\textsc{i} 
and \textsc{ii}
offset for \hdoneoh.
Figure~\ref{fewaveplot} illustrates the differences 
in the derived abundances between that study
and the present one.
As expected, the agreement is good for the optical lines
($\sigma =$~0.06~dex from 122~Fe~\textsc{i} lines;
 $\sigma =$~0.07~dex from 9~Fe~\textsc{ii} lines).
The zeropoint offset ($+$0.09~dex for Fe~\textsc{i};
$-$0.01~dex for Fe~\textsc{ii})
can be explained by the slightly warmer
\teff\ derived in the present study.
The dispersion increases considerably for the NUV lines near 2000~\AA\
($\sigma =$~0.39~dex from 14~Fe~\textsc{i} lines;
 $\sigma =$~0.22~dex from 9~Fe~\textsc{ii} lines).
Both sets of measurements show an offset in the abundances
derived from lines in the NUV, but the offset derived
by \citeauthor{roederer14d}\ 
(0.08~dex for Fe~\textsc{i}; 0.04~dex for Fe~\textsc{ii})
is smaller than that found here
(0.32~dex for Fe~\textsc{i}; 0.18~dex for Fe~\textsc{ii}).

To minimize any potential influences from non-LTE effects
on Fe~\textsc{i} lines,
we focus on the difference in Fe~\textsc{ii},
0.14~dex (i.e., 0.18 $-$ 0.04~dex).
The analysis presented in \citet{roederer14d} 
was performed by I.U.R.,
while the analysis presented here
was performed by H.R.J.\ and T.T.~
Both analyses used the same observed spectrum
and tools for analysis.
This 0.14~dex difference reflects the uncertainty that may be expected
when the analysis is performed by different individuals.

We test our underlying assumption that the 
phosphorus and iron abundances will track each other
if the continuous opacity in the NUV is altered.
We artificially adjust the NUV continuous opacity 
in our calculations so that the abundances derived from
NUV Fe~\textsc{i} and \textsc{ii} lines
agree with the abundances derived from
optical Fe~\textsc{i} and \textsc{ii} lines.
When using the adjusted opacities, the 
resulting [P/Fe] ratios change by less than 0.04~dex on average,
which is less than the statistical measurement uncertainty.
We conclude that our assumption is justified.

We apply a correction to our NUV abundances equal to the
mean offset derived from Fe~\textsc{ii} for the 
six stars in Table~\ref{offsettab}, 
$+$0.23~dex.
This correction is reflected in the abundances presented in
Table~\ref{phostab}.
We adopt the standard error of the mean (s.e.m.)\ 
of this correction,
0.03~dex,
as its statistical uncertainty.
We adopt the zeropoint uncertainty
in the optical-NUV abundance offset,
0.14~dex,
as the systematic uncertainty 
in this correction.

We emphasize that this is an empirical correction
designed to place abundances derived from lines in the NUV
on the same scale as abundances of numerous other
species derived from lines in the optical.
We explicitly assume that an unknown parameter(s)
is affecting the Fe~\textsc{ii} and P~\textsc{i} lines similarly.
Attempts to understand and account for the physical origin
of this offset are challenging and
beyond the scope of the present study.
The uncertainty in this empirical correction is
a substantial fraction of the total error budget,
so future investigations into its origin are 
greatly encouraged.

\subsection{Departures from LTE}
\label{nlte}

The P~\textsc{i} lines examined in Section~\ref{lines}
originate from four of the five lowest-lying
levels of the neutral phosphorus atom.
A simple LTE Boltzmann calculation
for the conditions found in the line-forming layers of
late-type stars
reveals that these levels comprise $\approx$~10--30\% of the 
population reservoir of neutral phosphorus.
The remaining $\approx$~70--90\% is found in the 
$^{4}S^{o}_{3/2}$ ground level.
The resonance transitions from this level are
found only in the vacuum UV.~
The $^{2}D^{o}_{5/2}$ level gives rise to
the 2136~\AA\ line, our primary abundance indicator,
and this level alone accounts for $\approx$~5--16\% of 
all neutral phosphorus in the line-forming layers.
For comparison, levels in 
the $^{4}P$ NIR Multiplet 1 employed by \citet{caffau11}
contain $<$~0.02\% of the neutral phosphorus
in the line-forming layers.

No non-LTE calculations exist for P~\textsc{i}
transitions under the conditions
found in late-type stellar atmospheres.
The relatively high percentage of neutral phosphorus
found in the $^{2}D^{o}_{5/2}$ level 
suggests that departures from LTE
are likely to be minimal.

\subsection{Statistical uncertainties}
\label{stat}

There are six sources of statistical uncertainty that
contribute to the random errors in our phosphorus abundance derivations.
Uncertainties in the \loggf\ values contribute 0.12~dex
(Section~\ref{loggfvalues}).
This term could be neglected in a
star-to-star differential analysis,
or when only a single line is used.
We include this term in the statistical uncertainty budget
because abundances in two stars 
(\mbox{BD$+$44~493} and \hdonefour)
are derived using lines other than the P~\textsc{i} 2136~\AA\ line.
Uncertainties in the damping constant for the 
P~\textsc{i} 2136~\AA\ line contribute
less than 0.05~dex
(Section~\ref{damping}).
The uncertainty in the mean offset between the 
optical and NUV abundance scales is 
0.03~dex
(Section~\ref{offset}).
The quadrature sum of these fixed sources of random uncertainty 
contributes 0.14~dex toward the statistical uncertainties.
We also include a term for the fitting uncertainty,
which is influenced by the S/N
and the line sensitivity to changes in abundance.
Finally, in computing the uncertainty in [P/Fe] reported
in Table~\ref{phostab}, we add these values in quadrature with
the random uncertainty in Fe~\textsc{i} (Table~\ref{elemtab})
and the uncertainties in [P/Fe] expected from
statistical uncertainties in the model atmosphere parameters
(Table~\ref{modelsigmatab}).

\subsection{Systematic uncertainties}
\label{sys}

We identify two sources of systematic uncertainties 
in the phosphorus abundance derivations.
One component quantifies the impact of the model atmosphere 
systematic uncertainties 
on the phosphorus abundances
(Section~\ref{analysis} and Table~\ref{modelsigmatab}).
The other component quantifies the zeropoint uncertainty
in the NUV abundance scale relative to the optical one,
0.14~dex
(Section~\ref{offset}).
These values and the total statistical uncertainty
(excluding the uncertainty in [P/Fe] from the model atmosphere
statistical uncertainties)
are used to compute the 
total uncertainty in $\log\epsilon$~(P) reported in 
Table~\ref{phostab}.

\subsection{Comparison with Caffau et al.\ (2011)}
\label{comparison}

Unfortunately, our sample does not include any
stars studied by \citet{caffau11},
so we cannot directly assess any systematic differences.
Both samples include several stars in the same metallicity range,
however.
If we assume that the stars in this region of metallicity
overlap ($-$0.7~$<$~[Fe/H]~$< -$0.1)
sample the same underlying stellar population,
we can evaluate whether any systematic differences may be present.
The mean [P/Fe] ratio for the nine stars from \citeauthor{caffau11}\
in this metallicity range is 
$\langle$[P/Fe]$\rangle = +$0.21~$\pm$~0.04~dex ($\sigma =$~0.12)
on our solar abundance scale.
In our sample, the five stars in this metallicity range have
$\langle$[P/Fe]$\rangle = -$0.03~$\pm$~0.08~dex ($\sigma =$~0.18)
using the abundances corrected for the optical-NUV offset.
These values are in 
marginal agreement at the $\approx$~2$\sigma$ level.
If the NIR and NUV P~\textsc{i} abundance indicators
are not severely affected by departures from LTE,
we assert that the $+$0.23~dex optical-NUV abundance offset
(Section~\ref{offset}) is 
necessary to avoid more significant differences
between our results and those of \citeauthor{caffau11} 

\section{Results and discussion}
\label{results}

\begin{figure*}
\begin{center}
\includegraphics[angle=0,width=3.1in]{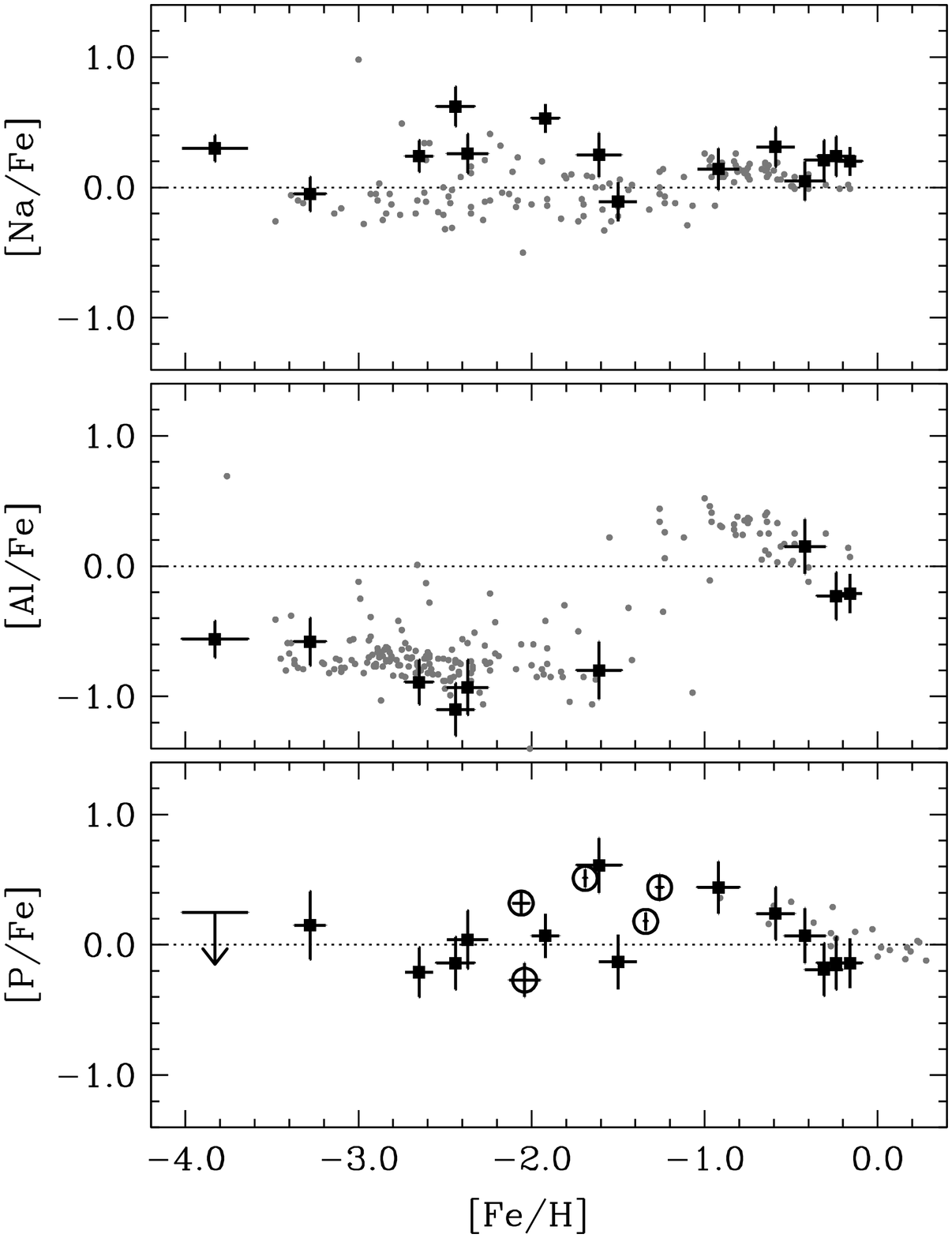}
\hspace*{0.1in}
\includegraphics[angle=0,width=3.1in]{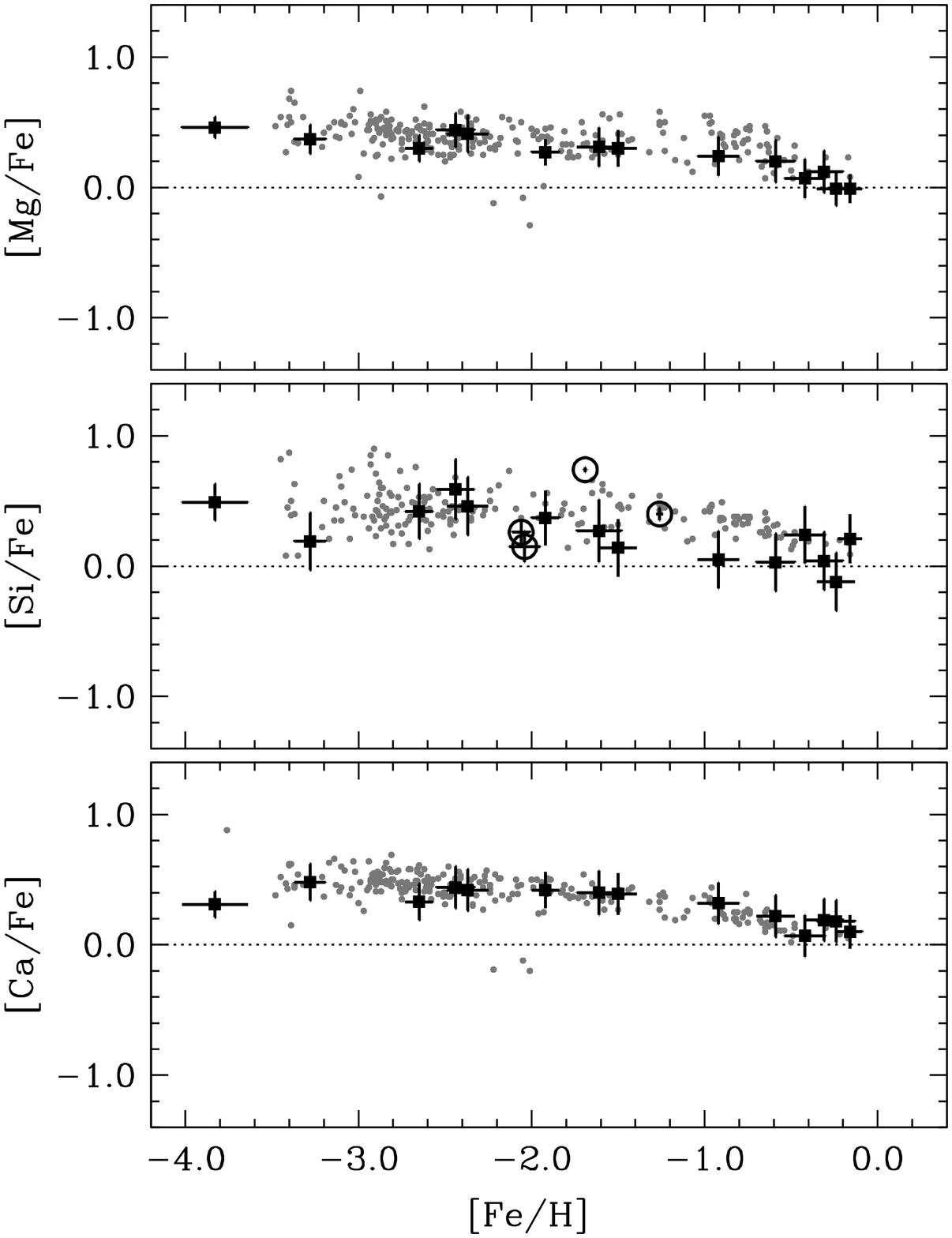}
\end{center}
\caption{
\label{abund1}
Derived abundance ratios.  
Stars in our sample are marked by large black symbols,
and stars in the comparison samples 
\citep{fulbright00,caffau11,roederer14c}
are marked by small gray symbols.
The open circles 
mark detections of Si~\textsc{ii} and
P~\textsc{ii} in DLA systems
(\citealt{outram99,molaro01,levshakov02,lopez02}; 
R.\ Cooke 2014, private communication).
The unpublished DLA reported by R.\ Cooke,
\mbox{J1558-0031}, has abundance ratios
[Fe/H]~$= -$2.06 and
[P/Fe]~$= +$0.32
in a cloud at $z_{\rm abs} =$~2.702;
see \citet{cooke14} for further details.
The dotted lines mark the solar ratios.
}
\end{figure*}

Our results are illustrated in Figure~\ref{abund1}.
The left-hand panel of Figure~\ref{abund1} shows the odd-$Z$
elements ([Na/Fe], [Al/Fe], and [P/Fe]), and 
the right-hand panel of Figure~\ref{abund1} shows the even-$Z$
elements ([Mg/Fe], [Si/Fe], and [Ca/Fe]).
Several sets of abundances derived by other studies
are shown for comparison.
The even-$Z$ $\alpha$-elements exhibit the familiar pattern
marked by a plateau at low metallicity and a gradual decline
to the solar ratios at [Fe/H]~$> -$1.0 or so.
The [Na/Fe] ratios scatter about the solar value.
The [Al/Fe] ratios are sub-solar at low metallicity
and solar or super-solar at high metallicity.
The derived sodium and aluminum abundances may be affected by 
departures from LTE
\citep{gehren04,andrievsky07,andrievsky08,lind11}.
We have made no attempt to correct for this in our data,
and only sodium in the \citet{roederer14c} comparison sample
has been corrected.
These corrections are negative and are large
enough to explain the offset between the
[Na/Fe] ratios in our study and those in the
\citeauthor{roederer14c}\ study at low metallicity.
Otherwise, the results for the 13~stars in our sample
with phosphorus detections
generally overlap with the comparison samples.

The weighted mean [P/Fe] for all 13~stars in our sample is
$+$0.04~$\pm$~0.07 ($\sigma =$~0.25).
The scatter about the mean, 0.25~dex, is 
comparable to the typical uncertainties in the 
[P/Fe] ratios, which range from 
0.17~dex to 0.26~dex.
A one-sided Kolmogorov-Smirnov test
of the null hypothesis that these 13~stars are drawn from
a normal distribution with a mean of 
$+$0.04
and standard deviation of 0.25~dex
returns a $p$-value of 0.48.
The null hypothesis 
cannot be excluded with any measure of confidence,
and there is no compelling evidence for
cosmic scatter in the [P/Fe] ratios.

Dividing the sample at [Fe/H]~$= -$1.0, however,
reveals possible differences
in the high- and low-metallicity groups of stars.
For the stars with [Fe/H]~$< -$1.0, excluding \bd,
we derive the weighted mean ratios as follows:\
$\langle$[P/Fe]$\rangle  = +$0.04~$\pm$~0.10 ($\sigma =$~0.28),
$\langle$[Na/Fe]$\rangle = +$0.23~$\pm$~0.06 ($\sigma =$~0.20),
$\langle$[Mg/Fe]$\rangle = +$0.34~$\pm$~0.02 ($\sigma =$~0.06),
$\langle$[Al/Fe]$\rangle = -$0.85~$\pm$~0.09 ($\sigma =$~0.19),
$\langle$[Si/Fe]$\rangle = +$0.35~$\pm$~0.06 ($\sigma =$~0.16), and
$\langle$[Ca/Fe]$\rangle = +$0.41~$\pm$~0.02 ($\sigma =$~0.05).
The mean [P/Fe] ratio in the low metallicity stars
is about 0.2~dex lower than that predicted
by one of the models by \citet{cescutti12},
[P/Fe]~$\approx +$0.2 to $+$0.3~dex,
based on their corrected yields for massive stars.
At high metallicity, [Fe/H]~$> -$1.0, 
a trend of decreasing [P/Fe] with increasing [Fe/H] 
may be present, and is given by
[P/Fe]~$= -$0.86($\pm$0.32)\,[Fe/H]~$-$~0.33($\pm$0.16).
A similar but shallower trend is also seen in the
\citet{caffau11} data at high metallicity,
[P/Fe]~$\approx -$0.4\,[Fe/H]~$+$~0.0.
Type~Ia supernovae are predicted to produce much smaller
quantities of $^{31}$P \citep{iwamoto99}, accounting for the [P/Fe] 
decrease at high metallicity.

Phosphorus' 
behavior is consistent with that of
a primary element in the stars in our sample,
in the sense that it may be synthesized from hydrogen
in proportions roughly equal to, e.g., the iron yield.
As shown in Figure~\ref{abund1},
this behavior is similar to that 
of the $\alpha$-elements,
with a 
constant
[P/Fe] at low metallicity.
This behavior is also found in several odd-$Z$ elements not found
on the $\alpha$ chain, including potassium.
Potassium has two stable isotopes, $^{39}$K and $^{41}$K,
and its solar abundance is dominated by $^{39}$K (93\%).
We have not derived potassium abundances for our sample;
however, an analysis of 38~subgiants with 
$-$3.0~$\leq$~[Fe/H]~$\leq -$1.0 by \citet{roederer14c}
finds $\langle$[K/Fe]$\rangle = +$0.30~$\pm$~0.02 ($\sigma =$~0.15).
The supernova models of \citet{woosley95}
predict that $^{31}$P and $^{39}$K
are both primary isotopes.
$^{31}$P is produced by neutron-capture on the products of the
oxygen- and neon-burning shells in their models,
and its ejection is not very sensitive to the explosion mechanism.
$^{39}$K is also produced by oxygen burning.
Our results 
support
predictions that phosphorus, like potassium,
may have
a primary origin at low metallicity.

Figure~\ref{abund1}
also shows the [Si/Fe] and [P/Fe] ratios
derived from detections of
Si~\textsc{ii} and P~\textsc{ii} 
in un- (or minimally-)depleted gas in four DLAs
\citep{outram99,molaro01,levshakov02,lopez02}.
These authors conclude that neither phosphorus nor iron
are significantly depleted from the gas phase 
on dust grains in these systems.
The absorbing clouds span redshifts 
2.33~$\leq z \leq$~3.39.
No abundance information is available for
sodium, magnesium, aluminum, or calcium in these DLAs.
Our stellar abundances overlap the range of the DLA abundances,
suggesting that the Galactic stars and DLAs 
could have been enriched by metals 
produced by similar nucleosynthesis channels
in the early Universe.

We defer more detailed comparisons 
of our data with supernova yield predictions 
and chemical evolution models to a 
companion paper \citep{jacobson14}.

\section{Summary and outlook}
\label{summary}

We have detected several P~\textsc{i} absorption lines
in the NUV spectra of late-type metal-poor stars.
All of these detections
are made using publicly-available STIS observations.
We have mainly used the P~\textsc{i} 2136.18~\AA\ line
to derive phosphorus abundances in 13~stars with 
$-$3.3~$<$~[Fe/H]~$< -$0.1.
The P~\textsc{i} 2553.26~\AA\ line
has been used to derive an upper limit in
one star with [Fe/H]~$= -$3.8.
Departures from LTE are likely to
have a negligible impact on the derived phosphorus abundances.
We also estimate the uncertainties resulting from an
offset between the optical and NUV abundance scales,
revealed by examination of Fe~\textsc{i} and Fe~\textsc{ii} lines,
whose cause is currently undetermined.

The derived [P/Fe] ratios 
compare well with previous work by \citet{caffau11},
who used several NIR P~\textsc{i} multiplets as abundance indicators.
Our results also overlap the range of [P/Fe] ratios found in five
undepleted DLA systems at high redshift.
The six stars in our sample with [Fe/H]~$> -$1.0
show a decreasing trend of [P/Fe] with increasing [Fe/H],
as \citeauthor{caffau11}\ also found.
The stars with [Fe/H]~$< -$1.0 show a 
constant [P/Fe] ratio, 
$+$0.04~$\pm$~0.10,
which is in 
marginal
agreement with
predictions made by the chemical evolution models of \citet{cescutti12}.

To improve the abundance precision,
the dominant sources of uncertainty must be addressed.
A significant fraction of 
the uncertainty in the \loggf\ value of this line
comes from the measurement uncertainty 
of the lifetime of its upper level (known to 13\%);
modern techniques routinely reach a precision of 5\% 
(e.g., \citealt{denhartog11})
or better
(e.g., \citealt{moehring06}).
Another large source of uncertainty is the 
offset between the optical and NUV abundance scales.
A differential analysis of the same
(set of) P~\textsc{i} line(s) in stars
with similar stellar parameters would minimize the
uncertainty in relative abundances.
Practical limitations
imposed by the stiff competition for \textit{HST} observing time
will limit future work to small numbers of bright stars,
but such observations provide the only 
demonstrated approach to studying 
the nucleosynthetic origins of phosphorus in the early Galaxy.

\acknowledgments

We wish to express our appreciation to the many investigators
whose observations have made this project possible.
We thank R.\ Cooke for providing an unpublished 
DLA abundance that appears in Figure~\ref{abund1}.
We also thank the referee for several excellent suggestions
that have improved this manuscript.
Some of the data presented in this paper 
were obtained from the Mikulski Archive for Space Telescopes (MAST),
the ESO Science Archive Facility,
and the Keck Observatory Archive.
STScI is operated by the 
Association of Universities for Research in Astronomy, Inc., 
under NASA contract NAS5-26555. 
This research has made use of NASA's 
Astrophysics Data System Bibliographic Services, 
the arXiv pre-print server operated by Cornell University, 
the SIMBAD and VizieR databases hosted by the
Strasbourg Astronomical Data Center, and
the ASD hosted by NIST.~
IRAF is distributed by the National Optical Astronomy Observatories,
which are operated by the Association of Universities for Research
in Astronomy, Inc., under cooperative agreement with the National
Science Foundation.
Generous support for Program AR-13246 has been 
provided by NASA through a
grant from the Space Telescope Science Institute, which is operated by the
Association of Universities for Research in Astronomy, Incorporated, under
NASA contract NAS~5-26555.
The work of T.~T.\ and E.~T.\ was
supported by the MIT UROP program.
A.~F.\ is supported by NSF CAREER grant AST-1255160.

 {\it Facilities:} 
\facility{ESO:3.6m (HARPS)},
\facility{HST (STIS)},
\facility{Keck:I (HIRES)},
\facility{Magellan~II (MIKE)},  
\facility{Smith (Tull)},
\facility{VLT:Kueyen (UVES)}


\begin{thebibliography}{}


\bibitem[Andrievsky et al.(2007)]{andrievsky07} Andrievsky, S.~M., 
Spite, M., Korotin, S.~A., et al.\ 2007, \aap, 464, 1081 

\bibitem[Andrievsky et al.(2008)]{andrievsky08} Andrievsky, S.~M., 
Spite, M., Korotin, S.~A., et al.\ 2008, \aap, 481, 481 

\bibitem[Anstee \& O'Mara(1995)]{anstee95} Anstee S.~D., O'Mara B.~J.\ 
1995, \mnras, 276, 859 

\bibitem[Asplund et al.(2006)]{asplund06} Asplund, M., Lambert, 
D.~L., Nissen, P.~E., Primas, F., \& Smith, V.~V.\ 2006, \apj, 644, 229 

\bibitem[Asplund et al.(2009)]{asplund09} Asplund, M., Grevesse, N., 
Sauval, A.~J., \& Scott, P.\ 2009, \araa, 47, 481 

\bibitem[Ayres(2010)]{ayers10} Ayres, T.~R.\ 2010, \apjs, 187, 149 

\bibitem[Barklem et al.(2000)]{barklem00} Barklem, P.~S., Piskunov, N., 
\& O'Mara, B.~J.\ 2000, \aaps, 142, 467 

\bibitem[Barklem \& Aspelund-Johansson(2005)]{barklem05} Barklem, P.~S., 
\& Aspelund-Johansson, J.\ 2005, \aap, 435, 373 

\bibitem[Bensby et al.(2014)]{bensby14} Bensby, T., Feltzing, S., \& 
Oey, M.~S.\ 2014, \aap, 562, A71 

\bibitem[Bernstein et al.(2003)]{bernstein03} Bernstein, R., 
Shectman, S.~A., Gunnels, S.~M., Mochnacki, S., 
\& Athey, A.~E.\ 2003, \procspie, 4841, 1694 

\bibitem[Bidelman(1960)]{bidelman60} Bidelman, W.~P.\ 1960, \pasp, 72, 24 

\bibitem[Caffau et al.(2005)]{caffau05} Caffau, E., Bonifacio, P., 
Faraggiana, R., et al.\ 2005, \aap, 441, 533 

\bibitem[Caffau et al.(2007)]{caffau07} Caffau, E., Steffen, M., 
Sbordone, L., Ludwig, H.-G., \& Bonifacio, P.\ 2007, \aap, 473, L9 

\bibitem[Caffau et al.(2010)]{caffau10} Caffau, E., Sbordone, 
L., Ludwig, H.-G., Bonifacio, P., 
\& Spite, M.\ 2010, Astronom.\ Nachr., 331, 725

\bibitem[Caffau et al.(2011)]{caffau11} Caffau, E., Bonifacio, P., 
Faraggiana, R., \& Steffen, M.\ 2011, \aap, 532, A98 

\bibitem[Castelli \& Kurucz(2003)]{castelli03} Castelli, F., \& Kurucz, R.~L.\
Proc.\ IAU Symp.\ No 210, Modelling of Stellar Atmospheres,
N.\ Piskunov et al., eds.\ 2003, A20

\bibitem[Castelli \& Hubrig(2004)]{castelli04} Castelli, F., \& 
Hubrig, S.\ 2004, \aap, 425, 263 

\bibitem[Cayrel et al.(2004)]{cayrel04} Cayrel, R., Depagne, E., Spite, M., 
et al.\ 2004, \aap, 416, 1117 

\bibitem[Cescutti et al.(2012)]{cescutti12} Cescutti, G., Matteucci, F., 
Caffau, E., \& Fran{\c c}ois, P.\ 2012, \aap, 540, A33 

\bibitem[Chen et al.(2002)]{chen02} Chen, Y.~Q., Nissen, P.~E., 
Zhao, G., \& Asplund, M.\ 2002, \aap, 390, 225 

\bibitem[Clegg et al.(1981)]{clegg81} Clegg, R.~E.~S., Tomkin, 
J., \& Lambert, D.~L.\ 1981, \apj, 250, 262 

\bibitem[Cooke et al.(2014)]{cooke14} Cooke, R., Pettini, M., \&
Jorgenson, R.~A.\ 2014, \apj, submitted (arXiv:1406.7003)

\bibitem[Cowley et al.(1969)]{cowley69} Cowley, C.~R., Elste, 
G.~H., \& Allen, R.~H.\ 1969, \apj, 158, 1177 

\bibitem[Crowther et al.(2002)]{crowther02} Crowther, P.~A., 
Hillier, D.~J., Evans, C.~J., et al.\ 2002, \apj, 579, 774 

\bibitem[Curtis et al.(1971)]{curtis71} Curtis, L.~J., 
Martinson, I., \& Buchta, R.\ 1971, \physscr, 3, 197 

\bibitem[Dekker et al.(2000)]{dekker00} Dekker, H., D'Odorico, 
S., Kaufer, A., Delabre, B., \& Kotzlowski, H.\ 2000, \procspie, 4008, 534 

\bibitem[Den Hartog et al.(2011)]{denhartog11} Den Hartog, E.~A., 
Lawler, J.~E., Sobeck, J.~S., Sneden, C., 
\& Cowan, J.~J.\ 2011, \apjs, 194, 35 

\bibitem[Ecuvillon et al.(2004)]{ecuvillon04} Ecuvillon, A., 
Israelian, G., Santos, N.~C., et al.\ 2004, \aap, 426, 619 

\bibitem[Edvardsson et al.(1993)]{edvardsson93} Edvardsson, B., 
Andersen, J., Gustafsson, B., et al.\ 1993, \aap, 275, 101 

\bibitem[Fran\c{c}ois(1987)]{francois87} Fran\c{c}ois, P.\ 1987, \aap, 176, 294 

\bibitem[Fran\c{c}ois(1988)]{francois88} Fran\c{c}ois, P.\ 1988, \aap, 195, 226 

\bibitem[Frebel et al.(2013)]{frebel13} Frebel, A., Casey, 
A.~R., Jacobson, H.~R., \& Yu, Q.\ 2013, \apj, 769, 57 

\bibitem[Friedman et al.(2000)]{friedman00} Friedman, S.~D., Howk, 
J.~C., Andersson, B.-G., et al.\ 2000, \apjl, 538, L39 

\bibitem[Fulbright(2000)]{fulbright00} Fulbright, J.~P.\ 2000, \aj, 120, 1841 

\bibitem[Gehren et al.(2004)]{gehren04} Gehren, T., Liang, Y.~C., 
Shi, J.~R., Zhang, H.~W., \& Zhao, G.\ 2004, \aap, 413, 1045 

\bibitem[Gray(1992)]{gray92} Gray, D.~F.\
1992, The observation and analysis of stellar photospheres, 2nd edition, 
Cambridge Univ.\ Press: Cambridge

\bibitem[Goldberg et al.(1960)]{goldberg60} Goldberg, L., Muller, 
E.~A., \& Aller, L.~H.\ 1960, \apjs, 5, 1 

\bibitem[Goswami \& Prantzos(2000)]{goswami00} Goswami, A., \& 
Prantzos, N.\ 2000, \aap, 359, 191 

\bibitem[Gratton \& Sneden(1988)]{gratton88} Gratton, R.~G., \& 
Sneden, C.\ 1988, \aap, 204, 193 

\bibitem[Guelin et al.(1990)]{guelin90} Guelin, M., Cernicharo, J., 
Paubert, G., \& Turner, B.~E.\ 1990, \aap, 230, L9 

\bibitem[Halfen et al.(2008)]{halfen08} Halfen, D.~T., 
Clouthier, D.~J., \& Ziurys, L.~M.\ 2008, \apjl, 677, L101 

\bibitem[Ito et al.(2009)]{ito09} Ito, H., Aoki, W., Honda, 
S., \& Beers, T.~C.\ 2009, \apjl, 698, L37 

\bibitem[Ito et al.(2013)]{ito13} Ito, H., Aoki, W., Beers, 
T.~C., et al.\ 2013, \apj, 773, 33 

\bibitem[Iwamoto et al.(1999)]{iwamoto99} Iwamoto, K., Brachwitz, 
F., Nomoto, K., et al.\ 1999, \apjs, 125, 439 

\bibitem[Jacobson et al.(2014)]{jacobson14} Jacobson, H.~R.,
Thanathibodee, T., Frebel, A., Cescutti, G., Matteucci, F.\
2014, \apjl, in press

\bibitem[Jenkins et al.(1986)]{jenkins86} Jenkins, E.~B., Savage, 
B.~D., \& Spitzer, L., Jr.\ 1986, \apj, 301, 355 

\bibitem[Johnson(2002)]{johnson02} Johnson, J.~A.\ 2002, \apjs, 139, 219 

\bibitem[J{\"o}nsson et al.(2011)]{jonsson11} J{\"o}nsson, H., 
Ryde, N., Nissen, P.~E., et al.\ 2011, \aap, 530, A144 

\bibitem[Junkkarinen et al.(1997)]{junkkarinen97} Junkkarinen, V., 
Beaver, E.~A., Burbidge, E.~M., et al.\ 1997, Mass Ejection from Active 
Galactic Nuclei, Astron.\ Soc.\ Pacific Conf.\ Ser.\ 128, ed.\ Arav et al.,
220 

\bibitem[Kato et al.(1996)]{kato96} Kato, K.-I., Watanabe, Y., 
\& Sadakane, K.\ 1996, \pasj, 48, 601 

\bibitem[Kimble et al.(1998)]{kimble98} Kimble, R.~A., Woodgate, 
B.~E., Bowers, C.~W., et al.\ 1998, \apjl, 492, L83 

\bibitem[Kobayashi et al.(2006)]{kobayashi06} Kobayashi, C., Umeda, 
H., Nomoto, K., Tominaga, N., \& Ohkubo, T.\ 2006, \apj, 653, 1145 

\bibitem[Kramida et al.(2013)]{kramida13} Kramida, A., Ralchenko, Yu., 
Reader, J., et al.\ 2013, NIST Atomic Spectral Database 
 (ver.\ 4.0--5.1) [online], available:\ 
http://physics.nist.gov/asd, 
National Institute of Standards and Technology, Gaithersburg, MD

\bibitem[Lambert \& Warner(1968)]{lambert68} Lambert, D.~L., \& 
Warner, B.\ 1968, \mnras, 138, 181 

\bibitem[Lawler et al.(2013)]{lawler13} Lawler, J.~E., Guzman, 
A., Wood, M.~P., Sneden, C., \& Cowan, J.~J.\ 2013, \apjs, 205, 11 

\bibitem[Lawrence(1967)]{lawrence67} Lawrence, G.~M.\ 1967, \apj, 148, 261 

\bibitem[Lebouteiller et al.(2006)]{lebouteiller06} Lebouteiller, V., 
Kunth, D., Lequeux, J., et al.\ 2006, \aap, 459, 161 

\bibitem[Lebouteiller et al.(2013)]{lebouteiller13} Lebouteiller, V., 
Heap, S., Hubeny, I., \& Kunth, D.\ 2013, \aap, 553, A16 

\bibitem[Leckrone et al.(1999)]{leckrone99} Leckrone, D.~S., 
Proffitt, C.~R., Wahlgren, G.~M., Johansson, S.~G., 
\& Brage, T.\ 1999, \aj, 117, 1454 

\bibitem[Lehner et al.(2003)]{lehner03} Lehner, N., Jenkins, 
E.~B., Gry, C., et al.\ 2003, \apj, 595, 858 

\bibitem[Levshakov et al.(2002)]{levshakov02} Levshakov, S.~A., 
Dessauges-Zavadsky, M., D'Odorico, S., \& Molaro, P.\ 2002, \apj, 565, 696 

\bibitem[Lind et al.(2011)]{lind11} Lind, K., Asplund, M., 
Barklem, P.~S., \& Belyaev, A.~K.\ 2011, \aap, 528, A103 

\bibitem[Lodders et al.(2009)]{lodders09} Lodders, K., Palme, H., 
\& Gail, H.-P.\ 2009, in Landolt B{\"o}rnstein, 
New Series, Vol.\ VI/4B, Tr\"{u}mper, ed., Springer-Verlag: Berlin
 (arXiv:0901.1149)

\bibitem[Lopez et al.(2002)]{lopez02} Lopez, S., Reimers, D., 
D'Odorico, S., \& Prochaska, J.~X.\ 2002, \aap, 385, 778 

\bibitem[Luck \& Bond(1981)]{luck81} Luck, R.~E., \& Bond, H.~E.\ 1981, 
\apj, 244, 919 

\bibitem[Magain(1987)]{magain87} Magain, P.\ 1987, \aap, 179, 176 

\bibitem[Marcolino et al.(2007)]{marcolino07} Marcolino, W.~L.~F., 
Hillier, D.~J., de Araujo, F.~X., \& Pereira, C.~B.\ 2007, \apj, 654, 1068 

\bibitem[Matrozis et al.(2013)]{matrozis13} Matrozis, E., Ryde, N., \& 
Dupree, A.~K.\ 2013, \aap, 559, A115 

\bibitem[Matteucci \& Fran\c{c}ois(1989)]{matteucci89} Matteucci, F., \& 
Fran\c{c}ois, P.\ 1989, \mnras, 239, 885 

\bibitem[Mayor et al.(2003)]{mayor03} Mayor, M., Pepe, F.,
Queloz, D., et al.\ 2003, The Messenger, 114, 20

\bibitem[Mel{\'e}ndez et al.(2009)]{melendez09} Mel{\'e}ndez, J., 
Asplund, M., Gustafsson, B., \& Yong, D.\ 2009, \apjl, 704, L66 

\bibitem[Milam et al.(2008)]{milam08} Milam, S.~N., Halfen, 
D.~T., Tenenbaum, E.~D., et al.\ 2008, \apj, 684, 618 

\bibitem[Moehring et al.(2006)]{moehring06} Moehring, D.~L., 
Blinov, B.~B., Gidley, D.~W., et al.\ 2006, \pra, 73, 023413 

\bibitem[Molaro et al.(2001)]{molaro01} Molaro, P., Levshakov, 
S.~A., D'Odorico, S., Bonifacio, P., 
\& Centuri{\'o}n, M.\ 2001, \apj, 549, 90 

\bibitem[Moore et al.(1934)]{moore34} Moore, C.~E., Babcock, 
H.~D., \& Kiess, C.~C.\ 1934, \apj, 80, 59 

\bibitem[Morton(1975)]{morton75} Morton, D.~C.\ 1975, \apj, 197, 85 

\bibitem[Nissen et al.(2004)]{nissen04} Nissen, P.~E., Chen, Y.~Q., 
Asplund, M., \& Pettini, M.\ 2004, \aap, 415, 993 

\bibitem[Nissen et al.(2007)]{nissen07} Nissen, P.~E., Akerman, C., 
Asplund, M., et al.\ 2007, \aap, 469, 319 

\bibitem[Outram et al.(1999)]{outram99} Outram, P.~J., Chaffee, 
F.~H., \& Carswell, R.~F.\ 1999, \mnras, 310, 289 

\bibitem[Peterson(1981)]{peterson81} Peterson, R.~C.\ 1981, \apj, 244, 989 

\bibitem[Piskunov \& Valenti(2002)]{piskunov02} Piskunov, N.~E., \& 
Valenti, J.~A.\ 2002, \aap, 385, 1095 

\bibitem[Placco et al.(2014)]{placco14} Placco, V.~M., Beers, T.~C.,
Roederer, I.~U., et al.\ 2014, \apj, 790, 34

\bibitem[Ram{\'{\i}}rez et al.(2013)]{ramirez13} Ram{\'{\i}}rez, 
I., Allende Prieto, C., \& Lambert, D.~L.\ 2013, \apj, 764, 78 

\bibitem[Reddy et al.(2003)]{reddy03} Reddy, B.~E., Tomkin, J., 
Lambert, D.~L., \& Allende Prieto, C.\ 2003, \mnras, 340, 304 

\bibitem[Reddy et al.(2006)]{reddy06} Reddy, B.~E., Lambert, 
D.~L., \& Allende Prieto, C.\ 2006, \mnras, 367, 1329 

\bibitem[Reiff et al.(2007)]{reiff07} Reiff, E., Jahn, D., 
Rauch, T., et al.\ 2007, 15th European Workshop on White Dwarfs, 
Astronom.\ Soc.\ Pacific Conf.\ Ser.\ 372, 237 

\bibitem[Roederer et al.(2010a)]{roederer10a} Roederer, I.~U., 
Sneden, C., Lawler, J.~E., \& Cowan, J.~J.\ 2010a, \apjl, 714, L123 

\bibitem[Roederer et al.(2010b)]{roederer10b} Roederer, I.~U., 
Sneden, C., Thompson, I.~B., Preston, G.~W., 
\& Shectman, S.~A.\ 2010b, \apj, 711, 573 

\bibitem[Roederer et al.(2012)]{roederer12d} Roederer, I.~U., 
Lawler, J.~E., Sobeck, J.~S., et al.\ 2012, \apjs, 203, 27 

\bibitem[Roederer et al.(2014a)]{roederer14a} Roederer, I.~U., 
Preston, G.~W., Thompson, I.~B., Shectman, S.~A., 
\& Sneden, C.\ 2014a, \apj, 784, 158 

\bibitem[Roederer et al.(2014b)]{roederer14c} Roederer, I.~U., 
Preston, G.~W., Thompson, I.~B., et al.\ 2014b, \aj, 147, 136 

\bibitem[Roederer et al.(2014c)]{roederer14d} Roederer, I.~U.,
Schatz, H., Lawler, J.~E., et al.\ 2014c, \apj, 791, 32

\bibitem[Ryan(1998)]{ryan98} Ryan, S.~G.\ 1998, \aap, 331, 1051 

\bibitem[Ryan et al.(1996)]{ryan96} Ryan, S.~G., Norris, 
J.~E., \& Beers, T.~C.\ 1996, \apj, 471, 254 

\bibitem[Ryde \& Lambert(2004)]{ryde04} Ryde, N., \& Lambert, D.~L.\ 
2004, \aap, 415, 559 

\bibitem[Samland(1998)]{samland98} Samland, M.\ 1998, \apj, 496, 155 

\bibitem[Sargent \& Searle(1967)]{sargent67} Sargent, W.~L.~W., \& 
Searle, L.\ 1967, \apjl, 150, L33 

\bibitem[Savage \& Lawrence(1966)]{savage66} Savage, B.~D., \& 
Lawrence, G.~M.\ 1966, \apj, 146, 940 

\bibitem[Savage \& Sembach(1996)]{savage96} Savage, B.~D., \& 
Sembach, K.~R.\ 1996, \apj, 470, 893 

\bibitem[Simmons \& Blackwell(1982)]{simmons82} Simmons, G.~J., \& 
Blackwell, D.~E.\ 1982, \aap, 112, 209 

\bibitem[Sneden(1973)]{sneden73} Sneden, C.~A.\ 1973, 
Ph.D.~Thesis, University of Texas at Austin

\bibitem[Sobeck et al.(2011)]{sobeck11} Sobeck, J.~S., Kraft, 
R.~P., Sneden, C., et al.\ 2011, \aj, 141, 175 

\bibitem[Spite et al.(2011)]{spite11} Spite, M., Caffau, E., 
Andrievsky, S.~M., et al.\ 2011, \aap, 528, A9 

\bibitem[Struve(1930)]{struve30} Struve, O.\ 1930, \apj, 71, 150 

\bibitem[Svendenius(1980)]{svendenius80} Svendenius, N.\ 1980, 
\physscr, 22, 240 

\bibitem[Takada-Hidai et al.(2002)]{takadahidai02} Takada-Hidai, M., 
Takeda, Y., Sato, S., et al.\ 2002, \apj, 573, 614 

\bibitem[Takada-Hidai et al.(2005)]{takadahidai05} Takada-Hidai, M., 
Saito, Y.-J., Takeda, Y., et al.\ 2005, \pasj, 57, 347 

\bibitem[Takeda \& Takada-Hidai(2011)]{takeda11} Takeda, Y., \& 
Takada-Hidai, M.\ 2011, \pasj, 63, 537 

\bibitem[Tenenbaum \& Ziurys(2008)]{tenenbaum08} Tenenbaum, E.~D., \& 
Ziurys, L.~M.\ 2008, \apjl, 680, L121 

\bibitem[Thielemann et al.(1996)]{thielemann96} Thielemann, F.-K., 
Nomoto, K., \& Hashimoto, M.-A.\ 1996, \apj, 460, 408 

\bibitem[Timmes et al.(1995)]{timmes95} Timmes, F.~X., Woosley, 
S.~E., \& Weaver, T.~A.\ 1995, \apjs, 98, 617 

\bibitem[Tominaga et al.(2007)]{tominaga07} Tominaga, N., Umeda, 
H., \& Nomoto, K.\ 2007, \apj, 660, 516 

\bibitem[Tomkin et al.(1985)]{tomkin85} Tomkin, J., Lambert, 
D.~L., \& Balachandran, S.\ 1985, \apj, 290, 289 

\bibitem[Tull et al.(1995)]{tull95} Tull, R.~G., MacQueen, 
P.~J., Sneden, C., \& Lambert, D.~L.\ 1995, \pasp, 107, 251 

\bibitem[Turner \& Bally(1987)]{turner87} Turner, B.~E., \& 
Bally, J.\ 1987, \apjl, 321, L75 

\bibitem[Uns\"{o}ld(1955)]{unsold55} Uns\"{o}ld, A., Physik der
Sternatmosph\"{a}ren, Springer-Verlag, Berlin, 1955, p.\ 332

\bibitem[Valenti \& Fischer(2005)]{valenti05} Valenti, J.~A., \& 
Fischer, D.~A.\ 2005, \apjs, 159, 141 

\bibitem[Vogt et al.(1994)]{vogt94} Vogt, S.~S., Allen, S.~L., 
Bigelow, B.~C., et al.\ 1994, \procspie, 2198, 362 

\bibitem[Vennes et al.(1996)]{vennes96} Vennes, S., Chayer, P., 
Hurwitz, M., \& Bowyer, S.\ 1996, \apj, 468, 898 

\bibitem[Wood et al.(2013)]{wood13} Wood, M.~P., Lawler, 
J.~E., Sneden, C., \& Cowan, J.~J.\ 2013, \apjs, 208, 27 

\bibitem[Wood et al.(2014)]{wood14} Wood, M.~P., Lawler, 
J.~E., Sneden, C., \& Cowan, J.~J.\ 2014, \apjs, 211, 20

\bibitem[Woodgate et al.(1998)]{woodgate98} Woodgate, B.~E., 
Kimble, R.~A., Bowers, C.~W., et al.\ 1998, \pasp, 110, 1183 

\bibitem[Woosley \& Weaver(1995)]{woosley95} Woosley, S.~E., \& 
Weaver, T.~A.\ 1995, \apjs, 101, 181 

\bibitem[Ziurys(1987)]{ziurys87} Ziurys, L.~M.\ 1987, \apjl, 321, L81 



\end{thebibliography}
\end{document}